\documentstyle[prl,aps,epsfig,multicol,amssymb,amsfonts]{revtex}
\renewcommand{\narrowtext}
{\begin{multicols}{2}\global\columnwidth20.5pc}
\renewcommand{\widetext}
{\end{multicols}\global\columnwidth42.5pc}
\def\top #1 {\vskip #1
\begin{picture}(290,80)(80,500)\thinlines\put(65,500)
{\line(1,0){255}}\put(320,500){\line(0,1){5}}\end{picture}}
\def\bottom #1 {\vskip #1
\begin{picture}(290,80)(80,500)\thinlines\put(330,500)
{\line(1,0){255}}\put(330,500){\line(0,-1){5}}\end{picture}}
\tighten
\begin{document}
\draft 
\title{Quasiclassical magnetotransport in a random array of antidots}
\author{D.~G.~Polyakov,$^{1,*}$ F.~Evers,$^1$
A.~D.~Mirlin,$^{1,2,\dagger}$ and P.~W\"olfle$^{1,2}$}
\address{$^1$Institut f\"ur Nanotechnologie, Forschungszentrum
Karlsruhe, 76021 Karlsruhe, Germany}
\address{$^2$Institut f\"ur Theorie der Kondensierten Materie,
Universit\"at Karlsruhe, 76128 Karlsruhe, Germany}
\maketitle
\begin{abstract}
We study theoretically the magnetoresistance $\rho_{xx}(B)$ of a
two-dimensional electron gas scattered by a random ensemble of
impenetrable discs in the presence of a long-range correlated random
potential. We believe that this model describes a high-mobility
semiconductor heterostructure with a random array of antidots. We show
that the interplay of scattering by the two types of disorder
generates new behavior of $\rho_{xx}(B)$ which is absent for only one
kind of disorder. We demonstrate that even a weak long-range disorder
becomes important with increasing $B$. In particular, although
$\rho_{xx}(B)$ vanishes in the limit of large $B$ when only one type
of disorder is present, we show that it keeps growing with increasing
$B$ in the antidot array in the presence of smooth disorder. The
reversal of the behavior of $\rho_{xx}(B)$ is due to a mutual
destruction of the quasiclassical localization induced by a strong
magnetic field: specifically, the adiabatic localization in the
long-range Gaussian disorder is washed out by the scattering on hard
discs, whereas the adiabatic drift and related percolation of
cyclotron orbits destroys the localization in the dilute system of
hard discs. For intermediate magnetic fields in a dilute antidot
array, we show the existence of a strong negative magnetoresistance,
which leads to a nonmonotonic dependence of $\rho_{xx}(B)$.
\end{abstract}

\vspace{8mm}

\narrowtext

\section{Introduction}
\label{intro}

In recent years, there has been a revival of interest in {\it
quasiclassical} transport properties of a two-dimensional electron gas
(2DEG). This has been largely motivated by the experimental progress
in controlled preparation of nanostructured semiconductor systems
\cite{ferry97} and, in particular, by the experimental and practical
importance of high-mobility heterostructures, in which impurities are
separated from the 2DEG by a wide spacer. On the theoretical side,
much of the recent interest in quasiclassics on the nanometer scale
has been related to the realization that the classical dynamics in a
disordered system is in fact much richer than the idealized Drude
picture suggests. Indeed, as far as ballistic mesoscopic systems are
concerned, electron transport has been studied in terms of
quasiclassical dynamics in great detail \cite{beenakker91}. However,
in diffusive systems with smooth disorder, a quasiclassical treatment
of electron kinetics is also appropriate and has been shown to lead to
new transport regimes. To describe the transport properties of such
system, one sometimes has to completely abandon theories based on the
Boltzmann equation. In Boltzmann transport theory, formulated in terms
of a set of relaxation times, quasiclassics leads to the Drude
results: analytical behavior of the ac conductivity $\sigma(\omega)$
at $\omega\to 0$, zero magnetoresistance (MR), etc. It has been
demonstrated, however, that quasiclassical {\it memory effects},
neglected in the conventional Boltzmann approach, yield a wealth of
anomalous transport properties of a 2DEG subject to {\it long-range}
disorder. In particular, non-Markovian kinetics gives rise to a
quasiclassical zero-frequency anomaly (see \cite{wilke00} and
references therein) in the ac response of a disordered 2DEG,
associated with return processes in the presence of smooth
inhomogeneities. Specifically, the return-induced correction to ${\rm
Re}\,\sigma (\omega)$ exhibits a kink $\propto |\omega|$. Another
manifestation of non-Markovian kinetics is a strong positive MR
\cite{mirlin99} in low magnetic fields, which is able to explain
\cite{evers99} the otherwise puzzling positive MR observed near
half-filling of the lowest Landau level in the fractional quantum Hall
regime. The strength of the above anomalies depends on the ratio
$d/l$, where $d$ is the correlation radius of disorder, $l$ the mean
free path, and grows with increasing $d/l$ as a power of this
parameter. Since quantum corrections are governed by a different small
parameter $1/k_Fl\ll 1$, where $k_F$ is the Fermi wavevector, it is
the long-range correlations of disorder with $k_Fd\gg 1$ that reveal
the quasiclassical anomalies. The condition $k_Fd\gg 1$ is typically
well satisfied in high-mobility semiconductor heterostructures.

In this paper, we consider the quasiclassical magnetotransport
properties of a 2DEG in a random array of antidots (AD). The transport
(dc and far-infrared) properties of AD arrays, both periodic and
random, have been the subject of many recent experiments, see, e.g.,
\cite{yevtushenko00,cina99,eroms99,nachtwei98,hochgrafe99,kukushkin97,luetjering96}
and references therein. In periodic arrays (for a review see
\cite{weiss97,fleischmann95}), interest has been focused on geometric
resonances which are associated with the periodicity and result, in
particular, in commensurability peaks in the MR
\cite{weiss91,schuster93,kang93,smet97}. On the other hand, random
arrays (see, e.g.,
\cite{gusev94,tsukagoshi95,luetjering96,nachtwei98,cina99,yevtushenko00})
constitute a remarkable disordered system where the ADs play the role
of hard-wall scatterers. We aim to study the MR in random AD arrays
and therefore assume that there exist two types of disorder: ADs,
which we model as impenetrable hard discs that scatter electrons, and
a smooth random potential, created in the heterostructures by charged
impurities behind a spacer. The quasiclassical MR in each of the
limits, where only one type of disorder is present, is well understood
by now (see Sec.~\ref{outline}). The purpose of the paper is to
demonstrate that the interplay of the two types of disorder yields new
physics that is absent in the limiting cases. We will show that,
although in the extreme of a strong magnetic field $B\to\infty$ the
dissipative resistivity $\rho_{xx}(B)$ {\it tends to zero} in either
of the limiting cases, it {\it diverges} in the presence of both types
of disorder. In particular, in the experimentally relevant situation
of relatively weak long-range disorder, i.e., when the mean free path
at zero $B$ is determined by scattering on ADs, the presence of the
weak long-range fluctuations will nonetheless become of crucial
importance with increasing $B$. It is worth noting that our model can
also be applicable to the description of the MR in an unstructured
2DEG with residual interface impurities playing the role of antidots
(large-angle scattering on residual impurities is known to become
important in unstructured samples with a wide spacer
\cite{coleridge91,saku96,umansky97}).

The paper is organized as follows. We give a brief review of past work
on the quasiclassical MR in Sec.~\ref{outline}. In the body of the
paper, we first consider in Sec.~\ref{body1} the MR at a moderately
strong $B$, when the collision time for scattering on ADs is not
affected by the magnetic field. Then, in Sec.~\ref{body2}, we turn to
the limit of strong $B$, where the collision time is renormalized as
compared to the Drude value and, in the extreme of very large $B$, a
single act of scattering involves ``skipping" of cyclotron orbits
along the surface of ADs. The whole picture turns out to be rather
complex and we choose the following logic of presentation. We fix the
zero-$B$ mean free paths for scattering on ADs and on the long-range
disorder and for different values of the density $n$ of ADs sweep the
magnetic field. In Sec.~\ref{body1} we start with the ``hydrodynamic
limit" of infinite $n$ and then gradually decrease $n$. In
Sec.~\ref{body2} we first consider a single act of scattering on an AD
for large $B$, then proceed to analyze the strong-$B$ transport in an
AD array. We present results of numerical simulations in
Sec.~\ref{numer} and summarize in Sec.~\ref{conclu} [where the
qualitative behavior of $\rho_{xx}(B)$ is illustrated in
Fig.~\ref{sketch}].

\section{Outline of known results: Limiting cases}
\label{outline}

\subsection{Lorentz model}
\label{outline1}

We start by briefly recalling the known results for the classical
Lorentz model in two dimensions (hard discs of radius $a$, randomly
placed with a concentration $n$; we assume that $na^2\ll 1$ and
$k_Fa\gg 1$, so that the mean free path $l_S=3/8na$). This is a good
model for an AD array in a heterostructure. As was pointed out in
\cite{baskin78,bobylev95}, Drude theory fails completely to describe
magnetotransport in this system. In the limit $n\to\infty$, $a\to 0$,
with $l_S$ held fixed, the resistivity depends on a single variable
$l_S/R_c$, where $R_c$ is the Larmor radius, and reads
\cite{bobylev95}, in units of the zero-$B$ resistivity $\rho_0$,
\begin{equation}
{\rho_{xx}(B)\over\rho_0}=F\left({l_S\over R_c}\right)~,
\label{e1}
\end{equation}
with $F(0)=1$ and $F(x\gg 1)\simeq 9\pi/8x\ll 1$. In Drude theory, the
dissipative resistivity is not affected by a magnetic field and
$F(x)=1$ for all $x$. The nontrivial kinetic problem (\ref{e1}) is in
fact fully solvable and the exact expression for the conductivity
tensor at arbitrary $l_S/R_c$ can be found in
\cite{bobylev95,kuzmany98} (see also \cite{kuzmany98,dmitriev01} for
numerical simulations of the problem).

\begin{figure}
\begin{center}
\vspace{-2.5cm}
\includegraphics[width=\columnwidth,clip]{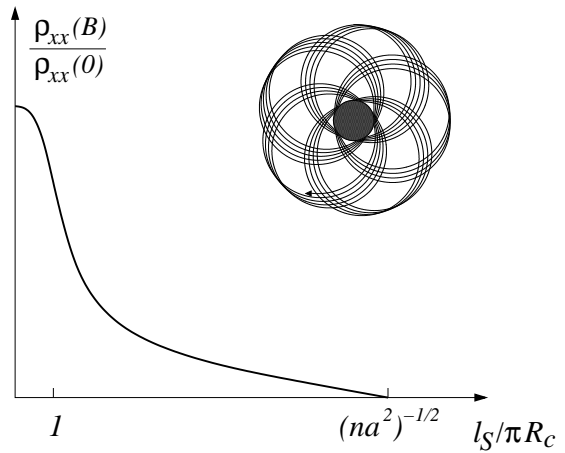}
\end{center}
\vspace{-2.5cm}
\caption{Schematic behavior of the magnetoresistivity $\rho_{xx}(B)$
as a function of $l_S/\pi R_c$ in the Lorentz model. Inset: Rosette
orbit of an electron bound to a hard disc (shown by the shaded circle)
in a magnetic field. }
\label{rosette}
\end{figure}

The falloff of $\rho_{xx}\propto B^{-1}$ is related to the peculiarity
of the Lorentz model: at finite $B$, there are electrons that move
freely in steady cyclotron orbits and never hit a scatterer; those
electrons do not contribute to $\rho_{xx}$ and their density grows
with increasing $B$. The conductivity is then due to electrons that
experience multiple collisions with a scatterer by moving in
``rosette'' orbits around it (Fig.~\ref{rosette}) until they hit
another scatterer, which results in a diffusive hopping of the
``rosette states". At finite concentration $n$, the Lorentz model has
a metal-insulator transition \cite{baskin78,bobylev95} at $R_c\sim
n^{-1/2}$: for larger $B$ the dissipative conductivity is strictly
zero, as shown in Fig.~\ref{rosette}.

\subsection{Long-range disorder}
\label{outline2}

Now let us recall what is known about MR in the case of a smooth
(allowing for a quasiclassical treatment) Gaussian (in the sense of
statistics of fluctuations) random scalar potential. There are two
sources of quasiclassical MR (we consider elastic scattering on an
isotropic Fermi surface). 

First, note that the MR is strictly zero in Boltzmann theory only in
the limit of white-noise disorder, whereas if disorder is correlated
on a finite spatial scale $d$, the collision-integral approximation
allows for a finite MR \cite{khveshchenko96,mirlin98}, due to a
cyclotron bending of trajectories within this correlation radius. This
simple effect is governed by the parameter $d/R_c$ [at small $B$ it
yields $\Delta\rho_{xx}/\rho_0\sim -(d/R_c)^2$].

\begin{figure}
\begin{center}
\includegraphics[width=0.8\columnwidth,clip]{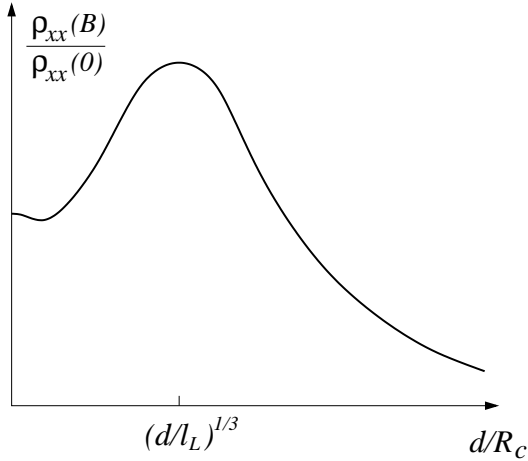}
\end{center}
\caption{Schematic behavior of the magnetoresistivity $\rho_{xx}(B)$
as a function of $d/R_c$ for a Gaussian smooth random potential. }
\label{lrange}
\end{figure}

Second, there is MR \cite{mirlin99} associated with memory effects
and, to calculate this, one has to go beyond the collision-integral
approximation. The memory effects are brought about by correlations of
scattering acts at the points where quasiclassical trajectories
self-intersect. These effects give the main contribution to
$\Delta\rho_{xx}/\rho_0$ at large enough $B$, where the governing
parameter is $d/\delta$ with $\delta$ being a characteristic shift of
the center of a cyclotron orbit after one revolution. For $R_c/d\agt
1$ the shift is (see \cite{fogler97} and references therein)
\begin{equation}
\delta\sim R_c(R_c/l_L)^{1/2}~,\quad R_c\agt d~,
\label{e2}
\end{equation}
where $l_L$ is the mean free path in the smooth random
potential (experimentally, $l_L/d\sim 10^2-10^3$ in high-mobility
samples). According to \cite{mirlin99}, $\Delta\rho_{xx}/\rho_0\sim
(d/\delta)^3\alt 1$. The return-induced contribution becomes much
larger than that related to the effect of $B$ on the collision
integral at $(\delta/R_c)^2\ll d/\delta$, i.e., at $R_c\ll
l_L(d/l_L)^{2/5}$. The exact expression for the MR in the limit
$(\delta/R_c)^2\ll d/\delta\ll 1$ in a heterostructure with a spacer
$d$ is \cite{mirlin99}
\begin{equation}
{\Delta\rho_{xx}(B)\over\rho_0}={2\zeta(3/2)\over\pi}
\left({d\over l_L}\right)^3\left({l_L\over R_c}\right)^{9/2}~.
\label{e3}
\end{equation}
This equation is valid with increasing magnetic field up to
$d/\delta\sim 1$, where $\Delta\rho_{xx}(B)/\rho_0$ becomes of order
unity. At higher fields, when $\delta/d\ll 1$, the strong positive MR
is followed by a sharp (exponential) falloff of $\rho_{xx}$ with
growing $B$ \cite{fogler97}:
\begin{equation} \ln \left({\rho_{xx}\over\rho_0}\right)\sim
-\left({d\over \delta}\right)^{2/3}~, 
\label{e4} 
\end{equation} 
which is due to the increasing adiabaticity of the electron dynamics
and the related quasiclassical localization \cite{fogler97,evers99}.
The self-intersection induced MR, given by Eq.~(\ref{e3}), may be
considered as a precursor of the adiabatic localization.  In the limit
of large $B$, when $R_c/d$ becomes small, $\delta$ is given by
\begin{equation}
\delta\sim R_c^2/(dl_L)^{1/2}~,\quad R_c\alt d~.
\label{e5}
\end{equation} 
The nonmonotonic behavior (\ref{e3}),(\ref{e4}) of the MR in the case
of a purely Gaussian long-range random potential is illustrated in
Fig.~\ref{lrange}.

To conclude the brief overview, it is worth noting that in the limit
of weak inhomogeneities the return-induced MR depends in an essential
way on the behavior of the disorder under time reversal. In
particular, it is strongly enhanced in the case of a random magnetic
field \cite{mirlin99}.

\section{Effect of cyclotron drift on transport in antidot arrays}
\label{body1}

\subsection{Parameters of the problem} 

We now turn to the MR in the presence of both a random array of hard
discs and long-range Gaussian disorder, which we characterize by the
mean free paths at zero magnetic field $l_S$ and $l_L$,
respectively. We assume that $l_S/l_L\ll 1$, which describes a typical
experimental situation. As in Sec.~\ref{outline}, $n$ will denote the
concentration of ADs, $a$ their radius, $d$ the correlation length of
the smooth random potential, $R_c$ the cyclotron radius, and $\delta$
the characteristic shift, due to scattering on the long-range
disorder, of the cyclotron orbit after one revolution. Throughout the
paper we assume $d/a\gg 1$.

We are interested in strong effects in the behavior of $\rho_{xx}(B)$:
for $l_S/l_L\ll 1$, these can only occur if $R_c/l_S\alt 1$. Moreover,
for the most part of the paper (namely, with the exception of
Sec.~\ref{scatt}), we consider magnetic fields which are sufficiently
strong in the sense that $\delta/d\alt 1$. In this case, the motion of
electrons is characterized by rapid cyclotron rotation around the
guiding center and slow drift of the latter along equipotential lines
of a smooth random potential. Most of these lines are closed, which
leads to localization of particles trapped on them. The effect of
scattering by ADs is to induce transitions between equipotential
contours and, in this way, allow the localized particles to escape.

\subsection{Hydrodynamic limit} 
\label{hydro}

Let us first consider the ``hydrodynamic limit" (\mbox{$n\to\infty$},
$a\to 0$, $l_S={\rm const}$). Clearly, in this limit, the effects
yielding the falloff of $\rho_{xx}\propto B^{-1}$ [Eq.~(\ref{e1})] are
washed out by infinitesimally weak long-range disorder. One might
think that then the Drude formula works and $\rho_{xx}(B)/\rho_0\simeq
1$ for all $B$. In fact, however, this is not true and even a small
($l_S/l_L\ll 1$) amount of smooth disorder becomes a relevant
perturbation with increasing $B$. Indeed, in the limit of large $B$
(namely, for $\delta/d\ll 1$), the problem can be mapped onto that of
advection-diffusion transport \cite{isichenko92}, i.e., of a Brownian
motion with a diffusion coefficient $D_0$ in a spatially random
velocity field ${\bf v}({\bf r})$ (``steady flow") with $\nabla\cdot
{\bf v}=0$ (``incompressible fluid"). In this mapping, the field ${\bf
v}({\bf r})$ describes the adiabatic drift of guiding centers of
cyclotron orbits due to long-range inhomogeneities and $D_0\sim
R_c^2/\tau_S$, where $\tau_S$ is the momentum relaxation time for
scattering on ADs. The result for the effective (macroscopic)
diffusion coefficient $D$ in the advection-diffusion problem
\cite{isichenko92} is
\begin{equation}
D\sim D_0(v_dd/D_0)^{10/13} 
\label{e6}
\end{equation}
if $v_dd\agt D_0$ and $D=D_0$ otherwise. Here $v_d$ is a
characteristic amplitude of the fluctuations of ${\bf v}({\bf r})$
[see Eqs.~(\ref{e11}),(\ref{e12}) below]. Hence the conductivity will
be strongly enhanced by even a weak long-range disorder provided
$v_dd/D_0\gg 1$. Since this parameter is a growing function of $B$,
the effect of smooth disorder is amplified by the magnetic field. The
reason is percolation of cyclotron orbits through long-range
inhomogeneities: the percolation-dominated $D$ can be written as a
product $v_dw$, where $w\ll d$ is a characteristic width of links of
the percolation network. The equation $w\sim d(D_0/v_dd)^{3/13}$ in
the advection-diffusion problem comes from the condition of
connectivity of the network $w^2v_d/L(w)\sim D_0$, where
\begin{equation}
L(w)\sim d(d/w)^{7/3}
\label{e7}
\end{equation}
is a typical length of the network link \cite{isichenko92}. Note that
the size $\xi(w)$ of the elementary cell of the percolation network
[i.e., a characteristic end-to-end distance for the link of length
$L(w)$] scales as \cite{isichenko92}
\begin{equation}
\xi(w)\sim d(d/w)^{4/3}~.
\label{e8}
\end{equation}

Although the advection-diffusion model has become popular for the
description of transport in the high-$B$ limit (in particular, in the
quantum Hall regime \cite{simon94,polyakov95}), we should be careful
to check if the scattering on ADs can actually be described in this
model in terms of the diffusion coefficient $D_0$. Clearly, this
requires that $w$ be larger than a hopping length for the diffusion
process, which means $w\gg R_c$. While this condition is satisfied in
the extreme of large $B$, a nontrivial transport regime may occur with
increasing $B$ in which $D\gg D_0$ but $w\ll R_c$ (as we will see
below, this is the case if the long-range disorder is not too
weak). In this regime, the main contribution to $D$ comes from
electrons that move freely along the critical links: the ``ballistic"
motion along the percolating path is contrasted with the transverse
(across the drift trajectory) diffusion in the advection-diffusion
regime. In other words, the number of collisions with ADs during the
drift along a critical link of the percolation network is now of order
unity. As in the advection-diffusion regime, the number of passages of
the network link between two consecutive changes of critical cells is
also of order unity. It follows that $w$ obeys the simple scaling
$L(w)\sim v_d\tau_S$, so that the result for $D\sim v_dw$ is
\begin{equation} 
D\sim v_dd(d/v_d\tau_S)^{3/7}~, 
\label{e9}
\end{equation} 
which should be compared, as in the advection-diffusion problem, with
$D_0$: Eq.~(\ref{e9}) is valid when $D\agt D_0$. Note that in this new
regime $D$ does not contain the hopping length, which may be even
larger than $d$.

We are now prepared to calculate the MR. The chaotic scattering on the
long-range potential crosses over into the adiabatic drift with
increasing $B$ at $R_c$ of order 
\begin{equation}
\tilde{R}_c=d(l_L/d)^{1/3}~, 
\label{e10}
\end{equation}
where $\delta/d$ becomes of order unity [cf.\
Eqs.~(\ref{e3}),(\ref{e4})]. At this field, the MR is still weak and
transport is completely determined by scattering on ADs, whereas at
larger $B$ we can already use the high-field formulas
(\ref{e6}),(\ref{e9}) and write $\rho_{xx}(B)/\rho_0\simeq D/D_0$. The
characteristic drift velocity $v_d$ that should be substituted into
Eqs.~(\ref{e6}),(\ref{e9}) reads (see, e.g., \cite{fogler97}): 
\begin{equation}
v_d=v_F{R_c\over (dl_L)^{1/2}}\,\,s\left({d\over R_c}\right)~,
\label{e11}
\end{equation}
where $v_F$ is the Fermi velocity and the function $s(x)$ is given by
\begin{equation}
s(x)\sim \left\{ \begin{array}{r@{\quad ,\quad}l}
x^{1/2} & x\ll 1 \\ 1 & x\gg 1 \end{array} \right. ~.
\label{e12}
\end{equation}
Which of Eqs.~(\ref{e6}),(\ref{e9}) should be used depends on the ratio
$w/R_c$, as explained above. Remarkably, the ratio $D/D_0$ for both
Eqs.~(\ref{e6}) and (\ref{e9}) depends only on two variables, $x=d/R_c$
and $v_d\tau_S/d=px^{-1}s(x)$, where
\begin{equation} p={l_S\over \sqrt{dl_L}}~. 
\label{e13} 
\end{equation}
As a result, the behavior of the MR as a function
of $B$ depends, at $a\to 0$, on the single parameter $p$,
so that we can write
\begin{equation}
{\rho_{xx}(B)\over \rho_0}=f\left({d\over R_c}\,,\,p\right)~.
\label{e14}
\end{equation}
Notice that even though we consider the case $l_S/l_L\ll 1$, the
parameter $p$ may be large since the long-range disorder is weak, i.e.,
$d/l_L\ll 1$.  

\begin{figure}
\begin{center}
\includegraphics[width=0.8\columnwidth,clip]{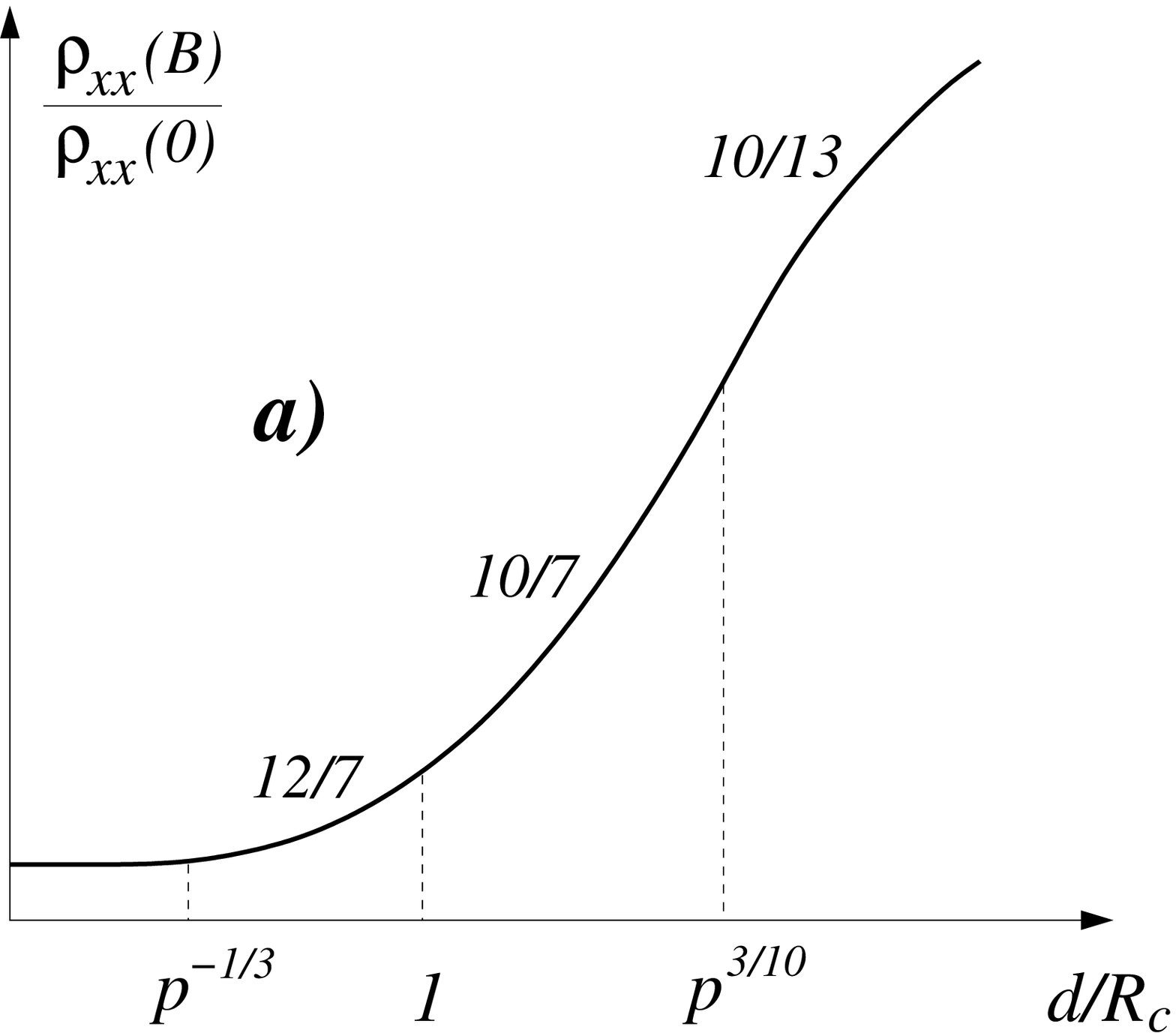}
\end{center}
\begin{center}
\includegraphics[width=0.8\columnwidth,clip]{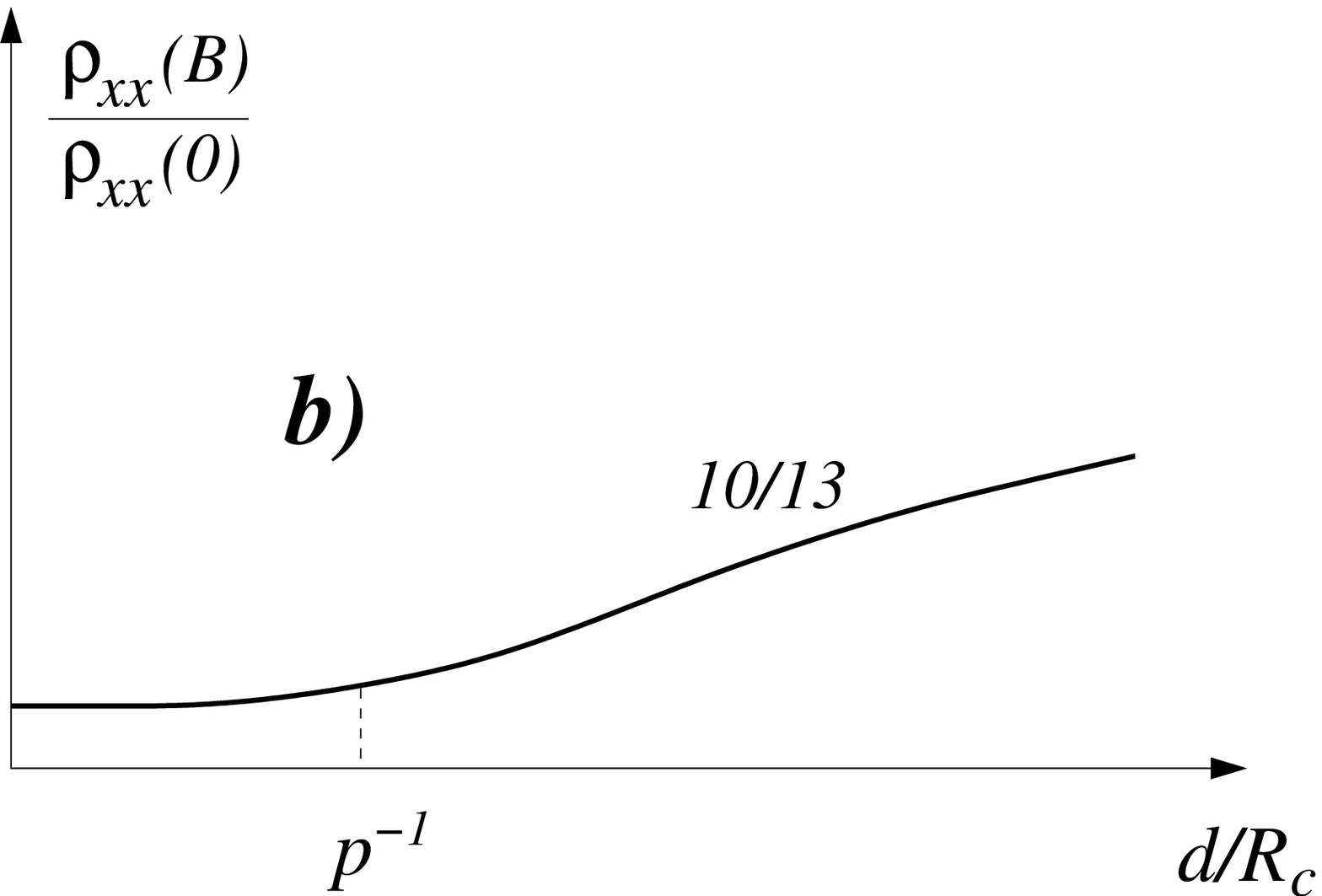}
\end{center}
\caption{Schematic behavior of the magnetoresistivity $\rho_{xx}(B)$
as a function of $d/R_c$ in the hydrodynamic limit $n\to\infty$, $a\to
0$, $l_S={\rm const}$ for a) $p=l_S/\sqrt{dl_L}\gg 1$ and b) $p\ll
1$. }
\label{hydrofig}
\end{figure}

At $p\gg 1$, the MR remains small, i.e., $f(x,p)\simeq 1$,
for all $x\ll p^{-1/3}$. On the upper boundary of this interval, the
percolation starts to renormalize $D$ in accordance with
Eq.~(\ref{e9}), which yields a power-law growth of $\rho_{xx}(B)$ with
further increasing $B$:
\begin{eqnarray}
f(x,p)\sim &&(px^3)^{4/7}~,\quad p^{-1/3}\ll x\ll 1~;\label{e15}\\
           &&(px^{5/2})^{4/7}~,\quad 1\ll x\ll p^{3/10}~.\label{e16}
\end{eqnarray} 
The scaling behavior changes between Eqs.~(\ref{e15}) and (\ref{e16}) at
$x\sim 1$ because of the change in the dependence of $v_d$ on $B$ at
$R_c/d\sim 1$. At still larger $B$, $D$ obeys Eq.~(\ref{e6}), which
gives
\begin{equation}
f(x,p)\sim (px)^{10/13}~,\quad p^{3/10}\ll x~.
\label{e17}
\end{equation}
Equations (\ref{e15})--(\ref{e17}) are illustrated in
Fig.~\ref{hydrofig}a.

At $p\ll 1$, the range of $x$ where the enhancement of the
conductivity is described by Eq.~(\ref{e9}) shrinks away, so that
$f(x,p)\simeq 1$ for all $x\ll p^{-1}$ and behaves according to
Eq.~(\ref{e17}) at larger $x$ (see Fig.~\ref{hydrofig}b). This
establishes the meaning of the parameter $p$: if $p\ll 1$, the Drude
regime does not match with increasing $B$ the advection-diffusion
regime (\ref{e6}) directly, but through the intermediate ``one-hop"
regime (\ref{e9}), whereas if $p$ is large, this intermediate phase is
absent. Equation (\ref{e17}) tells us that $\rho_{xx}(B)\propto
B^{10/13}$ at $B\to\infty$. The divergence takes place whatever the
ratio $l_S/l_L$, even if the long-range disorder is weak and does not
play a role at $B=0$. This behavior differs drastically from that
given by either of Eqs.~(\ref{e1}),(\ref{e4}).

\subsection{Finite density of antidots}
\label{finn}

So far, in Eqs.~(\ref{e14})--(\ref{e17}), the scattering on ADs has
been characterized by $l_S$ only, through the single parameter $p$
[Eq.~(\ref{e13})], which implies the hydrodynamic limit $n\to \infty$,
$a\to 0$. Now we take into account finite-$n$ effects. We begin as
before with the case of large $p$. New relevant dimensionless
parameters appear, in particular, $nR_cd$. Also, since for a fixed
$l_S$ decreasing $n$ means increasing $a$, the parameter $\delta/a$
may become relevant, in which case the scattering on ADs will be
affected by the magnetic field in an essential way and the collision
time will not be given by $\tau_S$. We will consider effects governed
by the parameter $\delta/a$ in Sec.~\ref{body2}. Until then, let us
assume that $\delta/a$ is sufficiently large, so that this parameter
plays no role for typical electron trajectories.

Clearly, if $\delta/a\gg 1$, stable rosette-states
\cite{baskin78,bobylev95} are still destroyed by the scattering on the
long-range potential. Naively, one could think that scattering on ADs
is then chaotic (no trace of the rosette-state dynamics) and
Eqs.~(\ref{e14})--(\ref{e17}) apply. In actual fact, provided
$nR_cd\ll 1$, multiple collisions with the same AD do occur even for
$\delta/a\gg 1$, as we will see below. At large $p$, the former
condition is satisfied with increasing $B$ before $\delta/a$ gets
small. Multiple returns in a dense AD array become possible because of
the adiabatic localization, which develops at $\delta/d\alt 1$.

Let us start by considering the drift regime under the condition
$nR_cd\ll 1$, $R_c/d\gg 1$. Typical trajectories of guiding centers
are closed loops of size $\sim d$, which means that trajectories of
electrons circling along cyclotron orbits are bound to within thin
rings of width $\sim d$ and radius $R_c$. The area of a strip between
the inner and outer radii of the rings is $\sim R_cd$ and, if
$nR_cd\ll 1$, in most rings there are no ADs. Electrons in these rings
are adiabatically localized and do not, in the adiabatic
approximation, contribute to $\rho_{xx}(B)$ (we will consider the
possibility of nonadiabatic decay of these states in
Sec.~\ref{nonad}). There are, however, rare rings with a single
AD. For electrons in these rings, a typical time $\tilde{\tau}_S$
between collisions with ADs is much shorter than $\tau_S$. Indeed, the
number of cyclotron revolutions before returning to the region of size
$\delta$ around the AD is typically $d/\delta$, while the probability
of hitting the AD during one such sweep is $\sim a/\delta$. It follows
that the number of cyclotron revolutions before the electron hits the
AD is $\sim d/a$, i.e.,
\begin{equation}
\tilde{\tau}_S\sim R_cd/v_Fa~, 
\label{e18} 
\end{equation} 
which gives $\tilde{\tau}_S/\tau_S\sim nR_cd\ll 1$. 

\begin{figure}
\begin{center}
\includegraphics[width=0.8\columnwidth,clip]{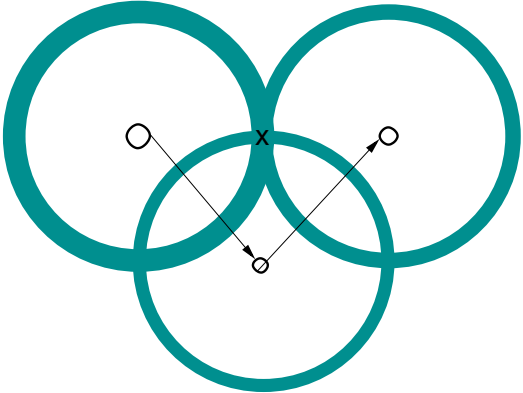}
\end{center}
\caption{A cartoon picture of scattering of a cyclotron orbit on an
antidot at $a\ll \delta\ll d\ll R_c\ll (nd)^{-1}$. The position of the
antidot is shown by a cross at the intersection of the rings (radius
$R_c$, characteristic width $d$) which represent the area ``covered" by
the drifting cyclotron orbit. The arrows denote hopping of the guiding
center of the orbit between drift trajectories shown by small
loops. Provided $nR_c^2\gg 1$, the particle will typically break away
from the antidot when it picks up a ring containing one more antidot
and hits the latter.}
\label{ringsfig}
\end{figure}

Now, a single collision with an AD does not lead to a breakaway from
the AD. In fact, the electron experiences multiple collisions with a
single AD and each time the center of the ring in which the electron
is circling hops a distance $\sim R_c$: one can visualize this process
as a random hopping of the center of the ring on a circle of
radius $R_c$ around the AD (see Fig.~\ref{ringsfig}). The electron
``sticks" to the AD for a time much longer than $\tilde\tau_S$. This
somewhat intricate dynamics of ``hopping rings" reminds the evolution
of the rosette states \cite{baskin78,bobylev95}: in effect it is the
adiabatic localization in the long-range potential that preserves the
character of the rosette-state dynamics. A breakup will eventually
happen when the electron picks up a very rare ring containing two ADs
(the existence of such rings in the area $R_c\times R_c$ implies that
$nR_c^2\gg 1$, which we assume in this derivation). It is
straightforward to check that the number of scatterings on a given AD
before getting to another one is $\sim 1/nR_cd\gg 1$. Multiplying the
latter by $\tilde\tau_S$, we find that the time it takes the electron
to change ADs (separated by a distance $\sim R_c$) is $\sim\tau_S$,
which yields the diffusion coefficient of electrons participating in
this type of transport $\sim D_0$, the same as in Drude theory. These
electrons, however, represent a small fraction of the total number of
electrons, namely $\sim nR_cd$. Hence, the contribution to the
macroscopic diffusion coefficient from drift orbits of a
characteristic size $d$ is $\sim D_0nR_cd\ll D_0$.

Having got the contribution of typical trajectories of size $d$, we
should take into account that upon hitting an AD the particle may hop
onto a drift trajectory of size $\xi$ larger than $d$. To put it
another way, although most trajectories with $\xi$ in the interval
$d\ll\xi\ll (nR_c)^{-1}$ are adiabatically localized, some of them hit
ADs and mix with the short-scale trajectories of size $d$ considered
above. This mixing increases the total fraction of delocalized
trajectories. To calculate the latter, note that the probability
density $P(\xi)$ for a point to belong to a drift trajectory of size
$\xi\agt d$ (we define $\xi$ as a characteristic radius of the area to
within which the trajectory is bounded) scales as
\begin{equation}
P(\xi)\sim d/\xi^2~.
\label{e19}
\end{equation} 
No critical exponents are involved here. One way to get a quick proof
of this is to realize that, according to percolation theory, for zero
altitude on a relief map of a random landscape, the number of contours
of radius $\xi$ in the area $\xi\times\xi$ is of order unity. It
follows that the (integrated over the altitude) fraction of space
occupied by contours of size $\sim\xi$ is $\sim Lw/\xi^2\sim d/\xi$,
where we used Eqs.~(\ref{e7}),(\ref{e8}), which yields
Eq.~(\ref{e19}).  Thus, the fraction of trajectories that are
delocalized due to collisions with ADs is evaluated by integration
\begin{equation}
\int_d^{R_c}\!\!
d\xi\,\,P(\xi)\,\,{\rm min}\{nR_c\xi, 1\}\sim nR_cd\,\ln(1/nR_cd)~,
\label{e20}
\end{equation}
which gives merely an additional logarithmic factor. One sees that the
characteristic $\xi\alt (nR_c)^{-1}$ are within the limits of
applicability of the derivation $\xi\ll R_c$ (trajectories with larger
$\xi$ give rise to a percolative contribution to $\rho_{xx}$
considered in Sec.~\ref{hydro}). Notice that drift trajectories that
do not hit ADs may be infinitely extended only with zero measure and
thus do not contribute to $D$. Accordingly, $D$ is evaluated as a
diffusion coefficient of electrons delocalized by the scattering on
ADs. Since the relevant $\xi\ll R_c$, the characteristic hopping
length associated with the change of ADs by these electrons is
$R_c$. The characteristic rate of hopping between two different ADs is
given by $n\left<\partial S/\partial t \right>$, where $\left<\partial
S/\partial t \right>\sim \int d\xi \,P(\xi)\, [A(\xi)/\tilde{\tau}_S
(\xi)]$ is the average rate at which the area explored by the particle
stuck to an AD grows in time. Here
\begin{equation}
\tilde{\tau}_S (\xi)\sim
\tilde{\tau}_S(d)\xi/d
\label{e21}
\end{equation} 
[with $\tilde{\tau}_S(d)$ defined in Eq.~(\ref{e18})] is the time the
particle resides on a trajectory of size $\xi$ before being scattered
out by the same AD, and $A(\xi)\sim R_c\xi$ is the area probed during
this time. These expressions for $\tilde{\tau}_S (\xi)$ and $A(\xi)$
are valid for $\xi\ll d(\delta/a)^{4/3}$, whereas at larger $\xi$ the
particle is scattered out before it comes full circle around the
closed trajectory and both quantities do not depend on $\xi$. We see
that $A(\xi)/\tilde{\tau}_S(\xi)$ does not depend on $\xi$ and
thus $\left<\partial S/\partial t \right>\sim v_Fa$ is determined by
$\xi\sim d$. It follows that the hopping rate does not change with the
inclusion of long trajectories and is given by
$\tau_S^{-1}$. Accordingly, the diffusion coefficient of delocalized
particles is $\sim D_0$. We finally get for the macroscopic diffusion
coefficient $D\sim D_0nR_cd\,\ln (1/nR_cd)$, or, for the MR:
\begin{equation} 
{\rho_{xx}(B)\over \rho_0}\sim nR_cd\,\,\ln {1\over
nR_cd}~.  
\label{e22} 
\end{equation} 
We thus see that, despite $\delta/a\gg 1$, the resistivity is strongly
suppressed as compared to the Drude result at $nR_cd\ll 1$. We will
return to this regime in Sec.~\ref{short}, where we will show that the
actual condition for Eq.~(\ref{e22}) to be valid is $\delta/a\gg
(nR_cd)^{-3/4}$, whereas at smaller $\delta/a$, $1\ll \delta/a\ll
(nR_cd)^{-3/4}$, a slight modification, namely in the logarithmic
factor in Eq.~(\ref{e22}), is necessary.

\begin{figure}
\begin{center}
\vspace{-2mm}
\includegraphics[width=0.8\columnwidth,clip]{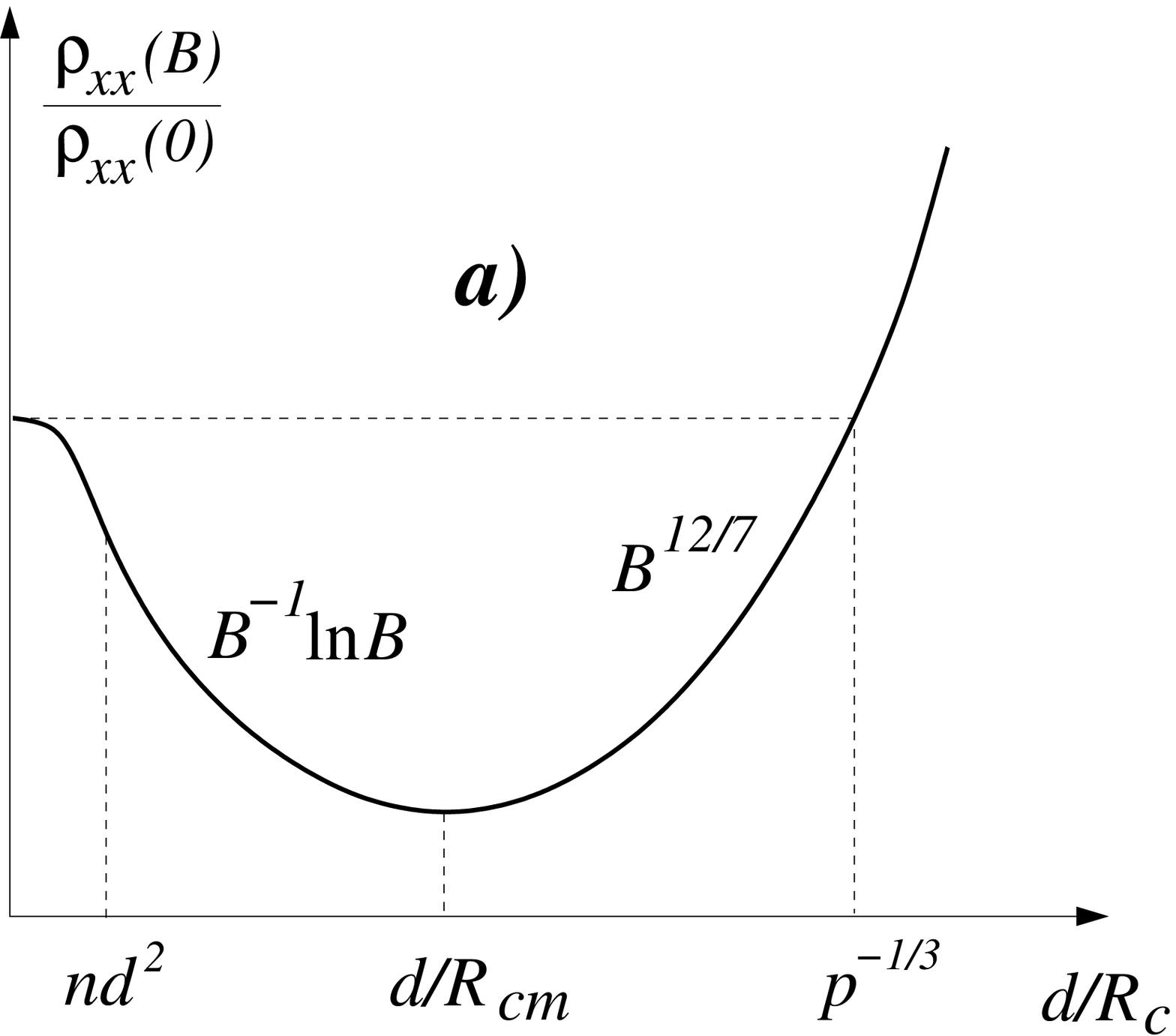}
\end{center}
\vspace{-5mm}
\begin{center}
\includegraphics[width=0.8\columnwidth,clip]{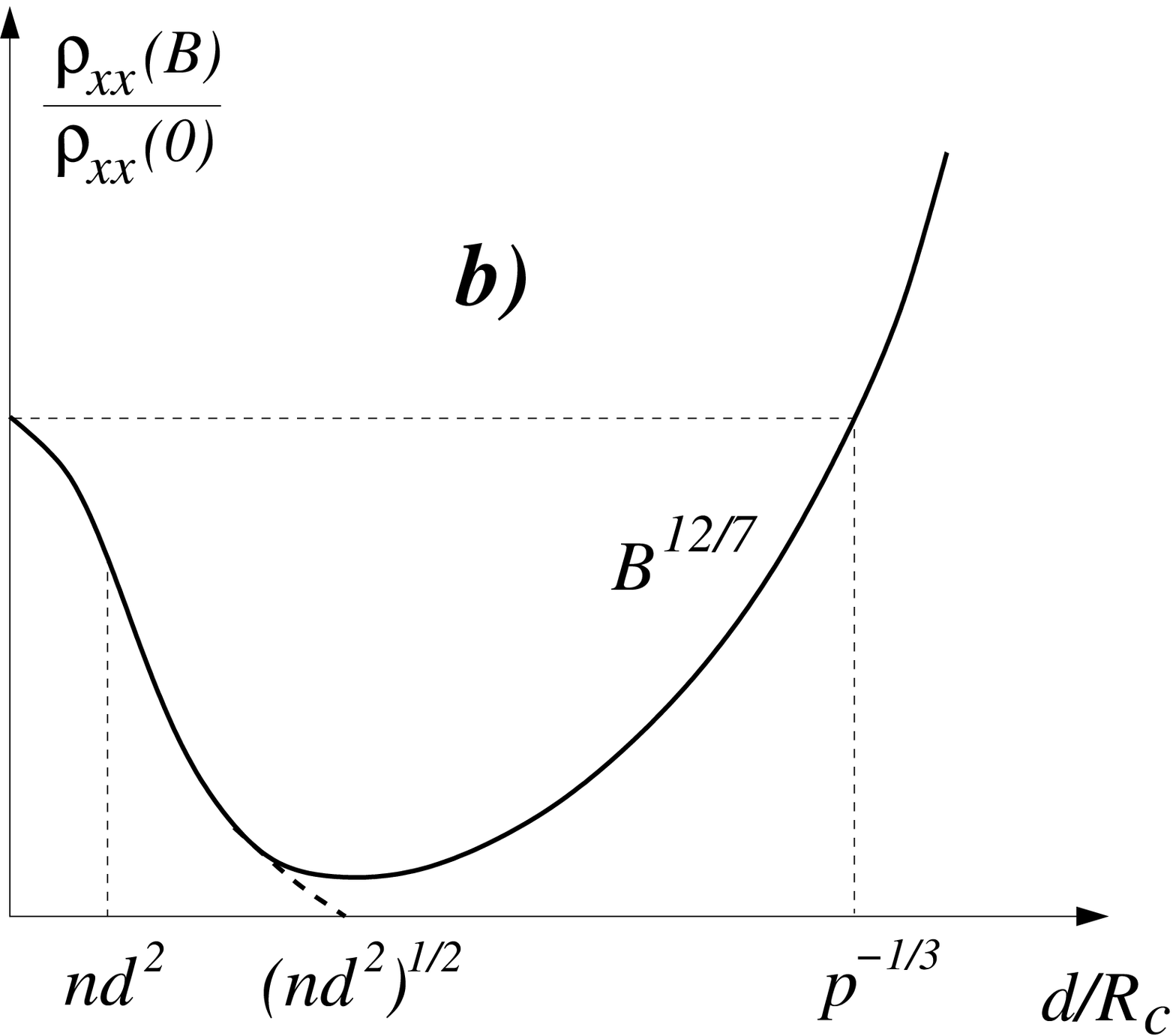}
\end{center}

\caption{Schematic behavior of $\rho_{xx}(B)$
as a function of $d/R_c$ for intermediate magnetic fields at large $p$
and $(d/l_L)^{1/3}\ll nd^2\ll p^{-1/3}$ for a) $d/R_{cm}\ll
(nd^2)^{1/2}$ and b) $d/R_{cm}\gg (nd^2)^{1/2}$.  The position of the
minimum of $\rho_{xx}(B)$ in Fig.~\ref{finite}a is given by
$d/R_{cm}\sim p^{-4/19}[nd^2\ln(1/nd^2p^{1/3})]^{7/19}$. Decreasing
$n$ leads to the nonmonotonic dependence of $\rho_{xx}(B)$ (cf.\
Fig.~\ref{hydrofig}, where $n\to\infty$). The dashed line in
Fig.~\ref{finite}b shows the behavior of $\rho_{xx}(B)$ given by
Eq.~(\ref{e23}). }
\label{finite}
\end{figure}

The falloff of the MR described by Eq.~(\ref{e22}) is illustrated in
Fig.~\ref{finite}a. Shown here is also the percolative growth of
$\rho_{xx}(B)$ into which the falloff crosses over at sufficiently
large $B$. As is clear from Fig.~\ref{finite}a, in the above
derivation we implicitly assumed that there exists a range of $B$
within which the percolative contribution is smaller than that given
by Eq.~(\ref{e22}). Indeed, since the percolative MR grows with
increasing $B$ (see Fig.~\ref{hydrofig}), $n$ should be small enough
for the $B^{-1}\ln B$ falloff not to be masked by the percolation. Let
us formulate the condition for the existence of the minimum in the
dependence of $\rho_{xx}(B)$ for $p\gg 1$. If one neglects (as
everywhere in this section) the nonadiabatic decay of drift
trajectories, the condition is $nd^2\ll p^{-1/3}$, as can be seen from
Fig.~\ref{finite}a. By matching Eqs.~(\ref{e22}) and (\ref{e15}) we
find that the minimum occurs at $R_c\sim R_{cm}$, where $d/R_{cm}\sim
p^{-4/19}[nd^2\ln(1/nd^2p^{1/3})]^{7/19}$. We will show in
Sec.~\ref{nonad}, by taking the nonadiabatic decay into account, that
the condition for the existence of the minimum actually reads
$\max\{nd^2,(d/l_L)^{1/3}\}\ll p^{-1/3}$, which means that
Fig.~\ref{finite}a correctly describes the MR if $(d/l_L)^{1/3}\ll
nd^2$.

To conclude this section, note that the interval of validity of
Eq.~(\ref{e22}) shrinks to zero if $R_c/d\alt 1$ and, therefore, the
above considerations describe the behavior of the MR only for $nd^2\ll
1$. According to Fig.~\ref{finite}, for large $p$ the nonmonotonic
behavior of $\rho_{xx}(B)$ develops for $nd^2\ll 1$. In the case of
small $p$ the picture is similar but the parameter $\delta/a$ becomes
relevant, which will be considered in Sec.~\ref{body2}.

\subsection{``Metal-insulator transition"}
\label{mit}

As noted above, the validity of the derivation of Eq.~(\ref{e22})
requires that $nR_c^2\gg 1$. This condition appears already in the
Lorentz model: if the opposite inequality is satisfied, the system
without long-range inhomogeneities would be insulating
($\rho_{xx}=0$). In the presence of long-range disorder, the
percolation mechanism of transport prevents $\rho_{xx}$ from vanishing
even at small $nR_c^2$. However, the parameter $nR_c^2$ determines the
interval of $B$ where the nontrivial mechanism of diffusion that leads
to Eq.~(\ref{e22}) is operative, namely $1\ll nR_c^2\ll R_c/d$. At
$nR_c^2\alt 1$, this mechanism is switched off in a manner inherent in
a continuous phase transition by formation of disconnected clusters of
trajectories, very much similar to the metal-insulator transition in
the Lorentz model \cite{baskin78,bobylev95}. Hence, the ``short-scale"
(as opposed to the percolative) MR [Eq.~(\ref{e22})] behaves near the
transition according to
\begin{equation}
{\rho_{xx}(B)\over \rho_0}=(nd^2)^{1/2}\ln{1\over nd^2}\,\,G
\left({B_c-B\over B_c}\right)~,
\label{e23}
\end{equation}
where $G(x)\sim x^t$ vanishes as a power law at $x\to 0$ on the
conducting side. The critical point $B=B_c$ corresponds to the
critical concentration $n=n_c\sim R_c^{-1/2}$. This behavior of
$\rho_{xx}(B)$ is shown in Fig.~\ref{finite}b. Comparing
Figs.~\ref{finite}a and \ref{finite}b, we see that the critical
falloff (\ref{e23}) is not masked by the percolation provided
$(nd^2)^{1/2}\ll d/R_{cm}$. If this condition is satisfied, the
minimum in the dependence of $\rho_{xx}(B)$ occurs at $nR_c^2\sim 1$.

We conjecture that the exponent $t$ in the function $G(x)$ can be
found by mapping the problem of percolation of skipping cyclotron
orbits onto that of percolation of the electric current through an
ensemble of conducting circles of radius $R_c$ scattered randomly with
the density $n$ (note that two rosette orbits of radius $2R_c$ do not
mix with each other if the distance between the centers of the
rosettes exceeds $2R_c$). The latter problem belongs to the
universality class of a two-dimensional percolation with a finite
threshold, for which many critical exponents are known (see, e.g.,
\cite{shklovskii84}; it is worth noting that the percolation of drift
trajectories considered above does not belong to this class). In
particular, the fraction of space occupied by the infinite cluster of
connected circles vanishes near the threshold as $(n-n_c)^\beta$ with
$\beta\simeq 0.14$, whereas the conductivity through the infinite
cluster exhibits a power-law behavior with the critical exponent
$t\sim 1.2$.

\subsection{Nonadiabatic decay}
\label{nonad}

In Sec.~\ref{finn}, we inferred the $B^{-1}\ln B$ falloff of
$\rho_{xx}(B)$ [Eq.~(\ref{e22})] by assuming that the drift picture is
applicable in the whole range $nR_cd\ll 1$. This is legitimate if
$n\tilde{R}_cd\gg 1$, where $\tilde{R}_c$ is defined in
Eq.~(\ref{e10}). Otherwise the Drude value of $\rho_{xx}=\rho_0$ holds
with increasing $B$ up to the field where the adiabatic dynamics
starts and there is an exponentially fast crossover between
$\rho_{xx}=\rho_0$ and $\rho_{xx}$ given by Eq.~(\ref{e22}), which is
governed by the nonadiabatic scattering.

\begin{figure}
\begin{center}
\includegraphics[width=0.8\columnwidth,clip]{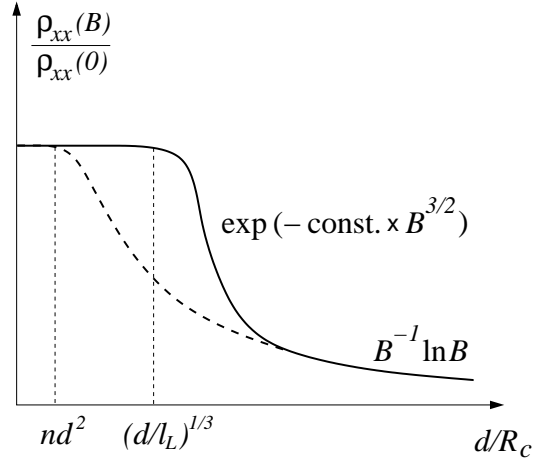}
\end{center}
\caption{Schematic representation of the behavior of $\rho_{xx}(B)$ at
$nd^2\ll (d/l_L)^{1/3}\ll p^{-1/3}\ll 1$ for moderately strong $B$
which correspond to the falloff of $\rho_{xx}(B)$ in
Figs.~\ref{finite}a,b. The thick dashed line shows $\rho_{xx}(B)$
without taking the nonadiabatic decay of drift trajectories into
account. Provided $nd^2\ll (d/l_L)^{1/3}$, the nonadiabatic
transitions stretch the range of $B$ where $\rho_{xx}(B)\simeq\rho_0$.
Beyond this range, $\rho_{xx}(B)$ falls off exponentially with
increasing $B$ until this falloff crosses over into the $B^{-1}\ln B$
behavior. }
\label{nonadfig}
\end{figure}

Let $D_{na}$ be the diffusion coefficient across drift trajectories
due to their nonadiabatic mixing. Since the rate of nonadiabatic
transitions depends exponentially on the parameter $d/\delta$ and is
therefore locally a wildly fluctuating quantity, we should be more
specific here: we define $D_{na}$ through a typical time $d^2/D_{na}$
that it takes to change two typical drift trajectories of size $\sim
d$ separated by a distance $\sim d$ (for thus defined diffusion
coefficient $\ln D_{na}$ scales as $d/\delta$, see, e.g.,
\cite{fogler97,evers99}). If $n\tilde{R}_cd\ll 1$, a particle which
initially resides on a ring with no AD will reach a ring containing
one in a time
\begin{equation} 
\tau_{na}\sim (d^2/D_{na})/nR_cd~.  
\label{e24}
\end{equation} 
Then there are two possibilities. If $\tilde{\tau}_S\ll d^2/D_{na}$
[with $\tilde{\tau}_S$ defined by Eq.~(\ref{e18})], the particle will
hit this AD, so that collisions with different ADs will occur at a
rate $\sim \tau_{na}^{-1}$, which yields a contribution to the
macroscopic diffusion coefficient $D\sim R_c^2/\tau_{na}\sim D_0
D_{na}\tilde{\tau}_S/d^2$, i.e.,
\begin{equation} D\sim D_0{D_{na}R_c\over v_Fda}~.  
\label{e25} 
\end{equation} 
Note that $D$ in this regime is proportional to a product of two
diffusion coefficients, $D_0$ and $D_{na}$. If, by contrast,
$\tilde{\tau}_S\gg d^2/D_{na}$, the particle will miss this AD and
will go on exploring phase space in a chaotic way, which gives
$D=D_0$. Since $d^2/D_{na}$ at $R_c\sim \tilde{R}_c$ is of the order
of the cyclotron frequency $v_F/R_c$, the crossover between these two
regimes takes place with increasing $B$ when $R_c\ll \tilde{R}_c$,
(logarithmically) deep in the drift regime. Hence, if
$n\tilde{R}_cd\ll 1$, the nonadiabatic decay of drift trajectories
stretches the region of a chaotic diffusion to the point at which
$D_{na}\tilde{\tau}_S/d^2\sim 1$, which occurs at $d/R_c$ only
logarithmically larger than $(d/l_L)^{1/3}$. At larger $B$, the
resistivity starts to fall off sharply, according to Eq.~(\ref{e25}),
until $D_{na}\tilde{\tau}_S/d^2$ becomes of order $nR_cd\ln
(1/nR_cd)$, where this exponential behavior crosses over into the
power-law falloff described by Eq.~(\ref{e22}). The characteristic
features in the behavior of $\rho_{xx}(B)$ at $p\gg 1$ associated with
the nonadiabatic decay of drift trajectories are illustrated in
Fig.~\ref{nonadfig} for the range of $B$ which corresponds to the
falloff of $\rho_{xx}(B)$ in Figs.~\ref{finite}a,b. At the point
$d/R_c\sim (d/l_L)^{1/3}$ shown in Fig.~\ref{nonadfig}, the adiabatic
localization starts to develop. 

Comparing Figs.~\ref{finite}a,b and \ref{nonadfig} one can formulate
the condition for the existence of the dip in the dependence of the MR
on $B$ for $p\gg 1$. If $nd^2\gg (d/l_L)^{1/3}$, the dip exists in the
case of $nd^2p^{1/3}\ll 1$, as in Figs.~\ref{finite}a,b. If, by
contrast, $nd^2\ll (d/l_L)^{1/3}$, the nonmonotonic behavior in
$\rho_{xx}(B)$ shows up for $dp/l_L\ll 1$. Figure~\ref{nonadfig}
illustrates the behavior of $\rho_{xx}(B)$ in the case when the
adiabaticity of motion in the long-range potential starts with
increasing $B$ well before the crossover to the percolative growth of
$\rho_{xx}(B)$. This is the case at $(d/l_L)^{1/3}\ll \min \{d/R_{cm},
(nd^2)^{1/2}\}$. It is clear, however, that the above considerations
of the nonadiabatic decay are equally valid for the case when the
opposite inequality is satisfied. The only difference is that, if
$(d/l_L)^{1/3}\gg \min \{d/R_{cm}, (nd^2)^{1/2}\}$, the exponential
falloff of $\rho_{xx}(B)$ crosses over into the percolative growth
directly, without passing through the intermediate $B^{-1}\ln B$
regime.

\section{Strong magnetic field: Renormalization of the collision time}
\label{body2}

Now let us turn to the extreme of strong $B$, where the parameter
$\delta/a$ becomes relevant. As will be shown below, the effect of
small $\delta/a$ is twofold. First, the time $\tau'_S$ between
collisions with two different ADs for particles drifting along
percolative trajectories (which is $\tau_S$ at large enough
$\delta/a$, how large---see below) gets longer. Second, a typical
hopping length $R_h$ across the drift trajectory due to the scattering
on an AD (which is $R_c$ at $\delta/a\gg 1$) gets smaller. Therefore,
before calculating the MR at small $\delta/a$ we have to consider how
the elementary scattering acts are modified. Below, Sec.~\ref{scatt}
deals with the scattering time $\tau'_S$, Secs.~\ref{skipor} and
\ref{nonadskip} with the hopping length $R_h$. Section \ref{rare}
studies the MR for small $\delta/a$.

\subsection{Effective scattering time}
\label{scatt}

Let us start by considering the case $\delta/a\ll 1$. To find
$\tau'_S$, notice that at $\delta/a\ll 1$ the length of the drift
trajectory between collisions $L_S\sim v_d\tau'_S$ cannot depend on
$a$, i.e., only the concentration $n$ matters. In the simplest case
$R_c\ll d$, the length $L_S$ clearly obeys the equation
\begin{equation}
nL_SR_c\sim 1~,
\label{e26}
\end{equation}
which is rewritten as
\begin{equation}
\tau'_S\sim\tau_S a/\delta~.
\label{e27}
\end{equation}
At $R_c\gg d$, however, there are different regimes for
$\tau'_S$. Namely, Eq.~(\ref{e27}) is valid for $\tau'_S$ at $R_c\gg d$
only as long as $L_S\ll d$, i.e., when the drift trajectory between
two collisions can be approximated as a straight line. If $L_S\gg d$,
which is the case at $nR_cd\ll 1$, the trajectory exhibits fractal
dimensionality on the scale of $L_S$. Specifically, the length of the
percolating trajectory $L_S$ and the distance $\xi_S$ from the
starting point are related to each other by 
$\xi_S\sim d (L_S/d)^{4/7}$ [see Eqs.~(\ref{e7}),(\ref{e8})], i.e.,
\begin{equation}
\xi_S\sim d(v_d\tau'_S/d)^{4/7}~.
\label{e28}
\end{equation}
It follows that the
cyclotron orbit passes many times through the same spatial regions,
which increases the time $\tau'_S$. Let first $L_S\gg d$ but $\xi_S\ll
R_c$ [i.e., $d\ll L_S\ll d(R_c/d)^{7/4}$]. In this regime, $\xi_S$ [in
contrast to $L_S$ in Eq.~(\ref{e26})] is of order $(nR_c)^{-1}$:
\begin{equation}
n\xi_S R_c\sim 1~,
\label{e29}
\end{equation}
which gives
\begin{equation}
\tau'_S\sim\tau_S\,{a\over \delta}\,{1\over (nR_cd)^{3/4}}~.
\label{e30}
\end{equation}
Now let $\xi_S\gg R_c$. To find $\tau'_S$ in this limit, one should
solve the following auxiliary problem. Collect all closed 
equipotential contours of size of order $\xi_S\gg d$ within the area
$\xi_S\times\xi_S$. They form a ``bundle" of width
\begin{equation}
w_S\sim d(d/v_d\tau'_S)^{3/7}\sim d(d/\xi_S)^{3/4}~.
\label{e31}
\end{equation}
The characteristic area covered by this bundle is $S(w_S)\sim L_Sw_S\ll
\xi_S^2$. Now enlarge the area by adding all points that are within a
distance $\Delta\agt w_S$ of the initial bundle. Doing so we get a new
bundle that occupies an area $S(\Delta)$. The question is what is
$S(\Delta)$ at $\Delta\sim R_c$ for $\xi_S\gg R_c\gg d$. Clearly,
$S(\Delta)\sim L_S\Delta$ as long as $\Delta\alt d$. However, at
$\Delta\gg d$, the area should exhibit a scaling behavior
$S(\Delta)\sim L_Sd(\Delta/d)^x$ with a nontrivial exponent $x$. To
find $x$, notice that at $\Delta\sim \xi_S$ we should have
$S(\xi_S)\sim \xi_S^2$. Using the equation $L_S\sim d(\xi_S/d)^{7/4}$
we thus get $x=1/4$. It follows that the area $S(R_c)$ scales at
$R_c\gg d$ as $L_Sd(R_c/d)^{1/4}$ [which can be represented as
$\xi_S^2(R_c/\xi_S)^{1/4}$ to see that most of space within the area
$\xi_S\times\xi_S$ is left empty]. Since $\delta\ll a$, it is clear
that if there is an AD in this area, it will be inevitably hit by the
cyclotron orbit. Therefore, the length $L_S$ obeys the equation
\begin{equation}
nS(R_c)\sim 1~, 
\label{e32}
\end{equation}
which yields $L_S\sim (nR_c)^{-1}(R_c/d)^{3/4}$ and
\begin{equation}
\tau'_S\sim \tau_S\,{a\over \delta}\,\left({R_c\over d}\right)^{3/4}~.
\label{e33}
\end{equation} 
We see that Eqs.~(\ref{e30}),(\ref{e33}) match each
other at $R_c\sim n^{-1/2}\gg d$. On the other hand,
Eqs.~(\ref{e27}),(\ref{e33}) match at $R_c\sim d$.

The above derivation of $\tau'_S$ at $\delta/a\ll 1$ shows that the
increase of the scattering time as compared to the ``Drude time"
$\tau_S$ is due to multiple passages of the cyclotron orbit through
the area $a\times a$. Clearly, at $R_c/d\ll 1$ no renormalization of
the scattering time occurs as long as $\delta/a\gg 1$. However, for
$R_c/d\gg 1$ the scattering time is renormalized with increasing $B$
already at some $\delta/a\gg 1$. Indeed, let $\delta/a\gg 1$ and
consider the case $L_S\gg d$, $\xi_S\ll R_c$. The drifting cyclotron
orbit experiences $L_S/\xi_S\gg 1$ returns to the area $d\times d$ and
each time it probes the fraction $a/\delta\ll 1$ of space within this
area. One sees that if the product of the two factors
$(L_S/\xi_S)(a/\delta)\gg 1$, then the collision time is much larger
than the Drude time $\tau_S$ and obeys Eq.~(\ref{e29}), which yields
Eq.~(\ref{e30}). Therefore, it is only when $\delta/a\gg
(nR_cd)^{-3/4}\gg 1$ that $\tau'_S=\tau_S$. Similarly, if $\xi_S\gg
R_c$, Eq.~(\ref{e33}) is valid for all $\delta/a\alt (R_c/d)^{3/4}$.

Inspection of Eqs.~(\ref{e27}),(\ref{e30}),(\ref{e33}) shows that the
dependence of $\tau'_S/\tau_S$ on $B$ is parametrized by two
parameters. In addition to $p$ [Eq.~(\ref{e13})], it is convenient to
introduce the parameter 
\begin{equation} 
\eta=nd^2p~, 
\label{e34}
\end{equation} 
which can be rewritten as $V_0d/\epsilon_Fa$, where $V_0$ is a
characteristic amplitude of fluctuations of the long-range potential,
$\epsilon_F$ the Fermi energy. The meaning of this parameter is that
it describes the position of the crossover $\delta/a\sim 1$ in terms
of the ratio $d/R_c$. Specifically, if $\eta\ll 1$, the crossover
occurs at $d/R_c\sim \eta^{2/3}$, whereas if $\eta\gg 1$, at
$d/R_c\sim \eta^{1/2}$. Note that the crossover point at $\eta\gg 1$
corresponds to $R_c\gg a$ [namely $R_c/a\sim (dl_L/a^2)^{1/4}$],
however large $\eta$ is. We thus have
\begin{equation} 
{\tau'_S\over \tau_S}=h\left({d\over R_c}\,,\,p\,,\,\eta\right)~.
\label{e35}
\end{equation} 
By changing $\eta$ with $p$ held constant, we change the concentration
$n$ at a fixed mean free path $l_S$. In Sec.~\ref{hydro}, we have
already studied the limit $\eta\to\infty$: in that case the scattering
time is not renormalized, so that $\tau'_S/\tau_S=1$ for all $B$
independently of $p$. However, Eqs.~(\ref{e27}),(\ref{e30}),(\ref{e33})
tell us that at any finite $\eta$ there exists a magnetic field above
which $\tau'_S/\tau_S$ starts to grow with increasing $B$. The
function $h(x,p,\eta)$ which describes this growth reads
\widetext \top{-2.7cm}
\begin{eqnarray}
h(x,p,\eta)\sim \left\{ \begin{array}{r@{\quad ,\quad}l}
x^2\eta^{-1} &  x\gg\max \{\eta^{1/2},1\}~; \\
x^{3/2}\eta^{-1} &  \eta^{2/3}\ll x\ll \min\{\eta p^{-1},1\}~,\quad  
p^3\ll\eta\ll 1~; \\
x^{9/4}p^{3/4}\eta^{-7/4} & \max\{\eta p^{-1},\eta^{7/9}p^{-1/3}\}\ll 
x \ll\eta^{1/2}p^{-1/2}~,\quad \eta\ll \min\{p,p^{-3/5}\}~; \\
x^{3/4}\eta^{-1} & \max\{\eta^{4/3},\eta^{1/2}p^{-1/2}\}\ll x\ll 1~,
\quad \eta\ll \min\{p,1\}~.
\end{array} \right.
\label{e36}
\end{eqnarray}
\bottom{-2.8cm}
\narrowtext \noindent

\begin{figure}
\begin{center}
\includegraphics[width=0.9\columnwidth,clip]{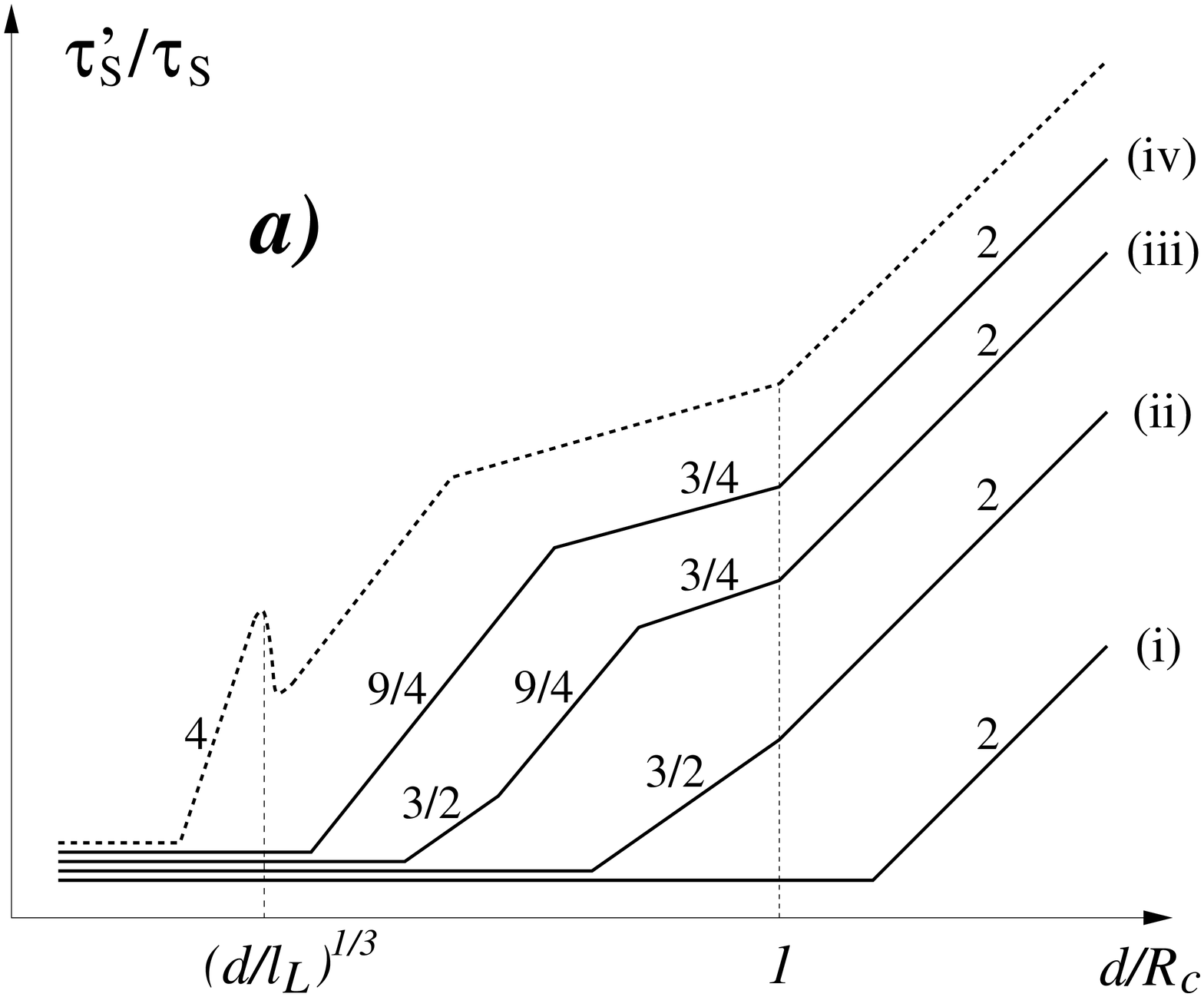}
\end{center}
\begin{center}
\includegraphics[width=0.9\columnwidth,clip]{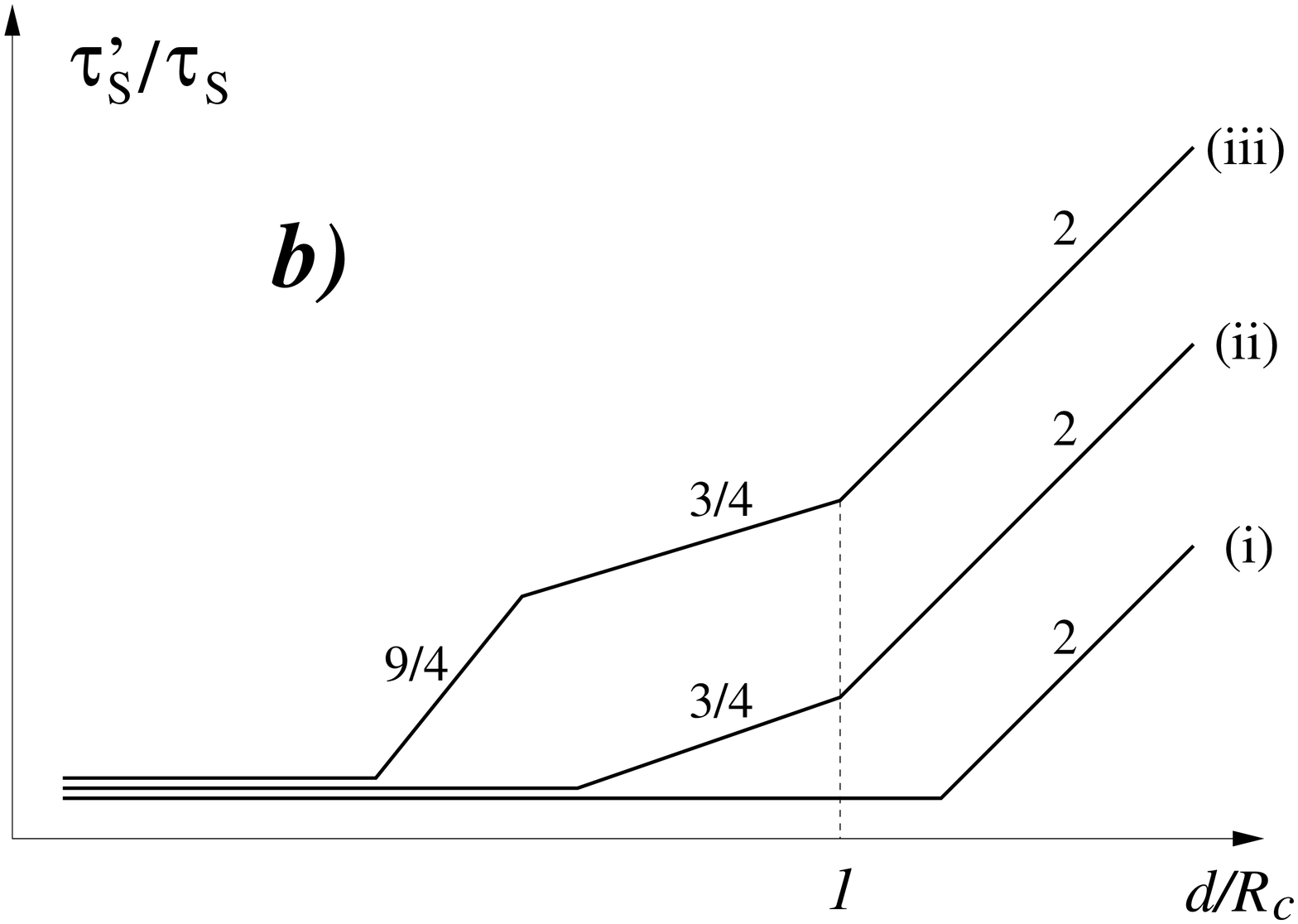}
\end{center}
\caption{Schematic behavior of $\tau'_S/\tau_S$ as a function of
$d/R_c$ on a log-log scale for a) $p\ll 1$ and b) $p\gg 1$. The numbers
denote the exponent of the power-law dependence of $\tau'_S/\tau_S$ on
$B$. Different curves in Fig.~\ref{taur}a illustrate how the
dependence of $\tau'_S/\tau_S$ on $B$ for a fixed $p$ is modified in
different ranges of $\eta$: (i) $\eta\gg 1$; (ii) $p\ll\eta\ll 1$;
(iii) $p^3\ll \eta\ll p$; (iv) $\eta\ll p^3$. Similarly in
Fig.~\ref{taur}b: (i) $\eta\gg 1$; (ii) $p^{-3/5}\ll \eta\ll 1$; (iii)
$\eta\ll p^{-3/5}$. The dashed line in Fig.~\ref{taur}a shows
the behavior of $\tau'_S/\tau_S$ for $\eta$ so small that the
renormalization of $\tau'_S/\tau_S$ starts with increasing $B$ already
in the diffusive regime (see the text). }
\label{taur}
\end{figure}

The behavior of $\tau'_S/\tau_S$ given by
Eqs.~(\ref{e35}),(\ref{e36}) is illustrated in Fig.~\ref{taur}. One
sees that in the limit of large $B$ the collision time grows as $B^2$,
whatever the parameters $p$ and $\eta$. If $\eta\gg 1$, which
corresponds to a sufficiently large concentration of ADs, this $B^2$
growth matches the Drude result directly [curves labeled by (i) in
Figs.~\ref{taur}a,b]. In a more dilute array of ADs, there appear
intermediate regimes, which proliferate as $\eta$ is decreased [curves
(ii)-(iv)]. Note that the smaller $\eta$ for a given $p$, the sooner
the renormalization of $\tau'_S$ starts with increasing $B$.

In Fig.~\ref{taur}a, we marked the point $d/R_c\sim (d/l_L)^{1/3}$
which corresponds to the crossover between the diffusion and drift
regimes [smaller and larger $d/R_c$ respectively, see
Eq.~(\ref{e10})]. Equations (\ref{e35}),(\ref{e36}) describe the drift
regime. However, as illustrated in Fig.~\ref{taur}a by the uppermost
(dashed) curve, if the concentration $n$ is sufficiently low, the
collision time is renormalized already in the diffusive
regime. Indeed, for small $n$, $\tau'_S/\tau_S$ calculated for
drifting electrons will be large at the crossover point to the
diffusive regime. A similar effect takes place for large $p$ as well
(not shown in Fig.~\ref{taur}b). In the diffusive regime,
$\tau'_S/\tau_S$ is still given by Eq.~(\ref{e29}), the only
difference is that now in place of Eq.~(\ref{e28}) one should take
\begin{equation}
\xi_S\sim \delta\,
(v_F\tau'_S/R_c)^{1/2}~,
\label{e37}
\end{equation}
which describes the diffusive motion of the
cyclotron orbit. Substituting this expression for $\xi_S$ in
Eq.~(\ref{e29}) yields
\begin{equation}
\tau'_S\sim\tau_S\,{l_L\over l_S}\,{1\over (nR_c^2)^2}~,
\label{e38} 
\end{equation} 
i.e., the collision time starts to grow as $B^4$ before entering the
drift regime, as shown in Fig.~\ref{taur}a by the dashed line. 
In the diffusive regime, $\rho_{xx}(B)$ is related to $\tau'_S$
by $\rho_{xx}(B)/\rho_0\sim\tau_S/\tau'_S$, which gives
\begin{equation}
{\rho_{xx}(B)\over\rho_0}\sim {l_S\over l_L}\,(nR_c^2)^2~.
\label{e39}
\end{equation} 
As follows from Eq.~(\ref{e38}), the crossover to the $B^4$ behavior
with increasing $B$ occurs at
\begin{equation}
d/R_c\sim
(nd^2)^{1/2}(l_S/l_L)^{1/4}~.
\label{e40}
\end{equation} 
Comparing Eq.~(\ref{e40}) with $(d/l_L)^{1/3}$ we conclude that the
crossover to the drift regime takes place at larger $B$ if $nd^2\ll
(d/l_L)^{2/3}(l_L/l_S)^{1/2}$. If, however, the concentration $n$ is
high enough, so that the opposite inequality is met, the region of
validity of Eqs.~(\ref{e38}),(\ref{e39}) shrinks away and no
renormalization of $\tau_S$ occurs in the diffusive regime.

Note a sharp change in the behavior of $\tau'_S/\tau_S$ at the
crossover between the drift and diffusion regimes
(Fig.~\ref{taur}a). The mismatch is due to the difference in the
fractal dimensionality of extended trajectories in the two regimes:
self-avoiding drift trajectories percolate in a super-diffusive way
and therefore explore the area faster. The sharp crossover has the
form of an exponential falloff of $\tau'_S/\tau_S$, governed by
rapidly developing adiabaticity of electron motion. If one
compares the times $\tau'_S$ obtained for the two (diffusion and
drift) regimes close to the crossover point $d/R_c\sim (d/l_L)^{1/3}$,
one can see that for small $n$, namely for $nd^2\ll
(d/l_L)^{5/7}(l_L/l_S)^{4/7}$, the ratio $\tau'_S/\tau_S\gg 1$ on both
sides. The dashed line in Fig.~\ref{taur}a illustrates just this
case. It is possible, however, that $n$ falls into the intermediate
range $(d/l_L)^{5/7}(l_L/l_S)^{4/7}\ll nd^2\ll
(d/l_L)^{2/3}(l_L/l_S)^{1/2}$, in which case $\tau'_S$ changes with
increasing $B$ in the following way: it first starts to grow in the
diffusive regime, then, after the crossover into the drift regime,
returns to the unrenormalized value $\tau_S$, and only with further
increasing $B$ begins to grow again.

At this point it is worth emphasizing once more that, while in the
diffusion regime all electrons behave in a similar way and $\tau'_S$
given by Eq.~(\ref{e38}) is characteristic to all electrons, upon
crossover into the drift regime the electrons find themselves divided
into different groups, characterized by different collision times (the
groups are mixed up only due to slow nonadiabatic
dynamics). Specifically, $\tau'_S$ in Eq.~(\ref{e30}) is the collision
time for electrons that move along extended (percolative) drift
trajectories. This time is renormalized even for $\delta\gg a$. On the
other hand, electrons that upon crossover to the drift regime find
themselves on typical drift trajectories of size $\sim d$ either do
not collide with ADs at all, and for them the collision time is
infinite (actually, provided they experience nonadiabatic transitions,
the collision time is finite but exponentially large, as shown in
Sec.~\ref{nonad}), or are characterized by the unrenormalized
collision time $\tau_S$ (at $\delta\gg a$), as explained in
Sec.~\ref{finn}. Hence, the renormalization of the collision time that
we have analyzed in this section will affect the percolative
contribution to the MR, which we will study in Sec.~\ref{rare}. For
$\delta\ll a$, the renormalization of $\tau_S$ will strongly affect
also the short-scale dynamics of electrons residing on typical
trajectories of size $d$, which will be considered in
Sec.~\ref{short}.

\begin{figure}
\begin{center}
\includegraphics[width=0.9\columnwidth,clip]{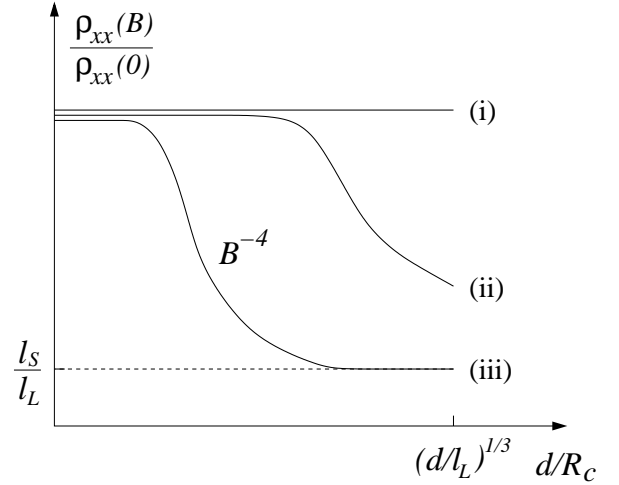}
\end{center}
\caption{Schematic behavior of $\rho_{xx}(B)$ as a function of $d/R_c$
in the diffusive regime $d/R_c\ll (d/l_L)^{1/3}$.  Different curves
illustrate the dependence of $\rho_{xx}(B)$ on $B$ in different ranges
of $n$: (i) $nd^2\gg (d/l_L)^{2/3}(l_L/l_S)^{1/2}$; (ii)
$(d/l_L)^{2/3}\ll nd^2\ll (d/l_L)^{2/3}(l_L/l_S)^{1/2}$; (iii)
$nd^2\ll (d/l_L)^{2/3}$.  }
\label{nmrdif}
\end{figure}

Now let us return to the diffusive regime. Clearly, Eq.~(\ref{e38})
is only valid as long as $v_F\tau'_S\ll l_L$, which is rewritten as
$nR_c^2\gg 1$. If the opposite limit, $nR_c^2\ll 1$, is realized with
increasing $B$ still in the diffusive regime [which is the case for a
very dilute array of ADs, namely for $nd^2\ll (d/l_L)^{2/3}$], in this
limit the scattering on ADs stops playing any role for diffusive
electrons. The collision time is then given by $l_L/v_F$ and
\begin{equation}
\rho_{xx}(B)/\rho_0=l_S/l_L
\label{e41}
\end{equation}
does not depend on $B$, whereas $\tau'_S/\tau_S$ keeps growing with
increasing $B$. Yet, the scattering on ADs will become relevant again
with further increasing $B$, once the system crosses over into the
drift regime (where the scattering on ADs will prevent the adiabatic
localization from completely suppressing the MR, as explained
above). This behavior corresponds to the case when the crossover to
the diffusive regime with decreasing $B$ occurs not in the region of
$B^{9/4}$ behavior, as shown in Fig.~\ref{taur}a, but in the $B^{3/4}$
region. The overall behavior of the MR in the diffusive regime is
illustrated in Fig.~\ref{nmrdif}. We consider the effect of the
renormalization of $\tau_S$ for diffusive electrons in detail
elsewhere \cite{short}.

As mentioned above, another effect of small $\delta/a$ is a
renormalization of the hopping length for the diffusive dynamics
across the drift trajectory due to the scattering on ADs, which we
will consider in more detail below. Since both the effective
scattering time and the hopping length are now modified, the shape of
the MR is no longer parametrized by a single parameter. Specifically,
we can replace $f$ in Eq.~(\ref{e14}) by a new function $g$ which is
given by
\begin{equation} 
g(x,p)\sim {\tau_S\over \tau'_S}\,f\left(x\,,\,p{\tau'_S\over
\tau_S}\right)~,
\label{e42}
\end{equation} 
in the ``one-hop" regime [cf.\ Eqs.~(\ref{e15}),(\ref{e16})], while in
the advection-diffusion regime [cf.\ Eq.~(\ref{e17})] we now have
\begin{equation} 
g(x,p)\sim {\tau_S\over \tau'_S}\left({R_h\over
R_c}\right)^{6/13}f\left(x\,,\,p{\tau'_S\over \tau_S}\right)~.
\label{e43}
\end{equation}
A crossover between the two regimes occurs at $v_d\tau'_S\sim L(R_h)$,
where $L(w)$ is defined in Eq.~(\ref{e7}). The ratio $\tau'_S/\tau_S$
is given by Eqs.~(\ref{e35}),(\ref{e36}). The ratio $R_h/R_c$ will
be calculated in Secs.~\ref{skipor},\ref{nonadskip}.

\subsection{Skipping orbits}
\label{skipor}

\begin{figure}
\begin{center}
\includegraphics[width=0.6\columnwidth,clip]{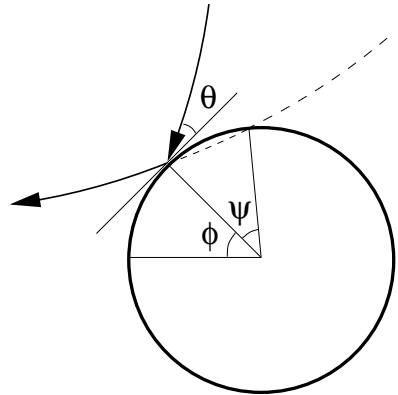}
\end{center}
\caption{Geometry of scattering of a skipping cyclotron orbit on an
antidot: $\theta$ is the angle of incidence, $\phi$ the polar angle of
the point at which the collision occurs, $\psi$ the incremental
increase of the angle $\phi$ between two consecutive collisions. }
\label{angles}
\end{figure}

In Sec.~\ref{scatt}, we derived general expressions for the
percolative MR for small $\delta/a$ in terms of $\tau'_S$ and $R_h$
and calculated the effective scattering time $\tau'_S$. Let us now
evaluate $R_h$. The scattering problem for a cyclotron orbit that
collides with a hard disc is nontrivial at $\delta/a\ll 1$. To begin
with, notice that at $\delta/a\ll 1$ the drifting cyclotron orbit
first hits the disc boundary at a small angle $\theta_1\ll 1$ (see
Fig.~\ref{angles}). A simple geometric consideration yields a
characteristic $\theta_1$ for a particle incident on the disc with a
drift-shift vector $\bbox{\delta}$ (i.e., with a drift velocity
${\bbox\delta}v_F/2\pi R_c$):
\begin{equation} 
\theta_1\sim \left(-{\bbox{\delta}\cdot {\bf e}\over
a}\right)^{1/2}\left({R_c+a\over R_c}\right)^{1/2}~, 
\label{e44}
\end{equation}
where $\bf e$ is the unit vector normal to the surface of the disc at
the point of the collision. One sees that the angle of incidence
vanishes when we let $\delta\to 0$, so that the skipping cyclotron
orbit in effect starts to ``roll over" the disc. Clearly, if it were
not for the drift during the rollover, the collision angle $\theta$
would be conserved, being the same each time the particle returns to the
disc after one cyclotron revolution. Therefore, although the drift is
slow, in the sense that the typical hopping length for the guiding
center after one collision $\sim R_c\theta\gg\delta$, it is because of
the drift during the skipping process that the particle eventually
breaks away from the AD. The length $R_h$ is then understood as a
shift of the guiding center at the point of the breakaway with respect
to the equipotential contour along which it was drifting right before the
first hit.

Since $\delta/a\ll 1$, we can treat the drift during the rollover
perturbatively. The system of equations that describe the skipping 
in the absence of drift is given by
\begin{eqnarray}
\phi_{n+1}=&&\phi_n+\psi (\theta_n)~; \label{e45}\\
\theta_{n+1}=&&\theta_n~, \label{e46}
\end{eqnarray}
where $\phi_n$ is the polar angle (along the surface of the disc)
defining the point of the $n$-th collision, $\theta_n$ the collision
angle, see Fig.~\ref{angles}. The function $\psi(\theta)$ obeys the
equation $R_c\sin (\theta -\psi/2)=a\sin (\psi/2)$, which reduces to
the linear relation
\begin{equation}
\psi(\theta)={2R_c\over R_c+a}\,\,\theta
\label{e47}
\end{equation}
in the limit of small $\theta$. Equation (\ref{e46}) says $\theta$ is
the integral of motion for the skipping process without drift. In the
presence of drift, $\theta_n$ acquires an $n$-dependent correction
$\Delta\theta_n=\theta_{n+1}-\theta_n$. To first order in $\delta$ the
correction reads
\begin{equation}
\Delta\theta_n\simeq -\,{R_c+a\over R_ca}\,\,{{\bbox\delta}_n{\bf
e}(\phi_n)\over \theta_n}~,
\label{e48}
\end{equation}
where ${\bbox\delta}_n$ is the drift shift between the $n$-th
and $(n+1)$-st collisions, ${\bf e}(\phi_n)$ the unit vector
perpendicular to the surface of the disc at the point of the $n$-th
collision. Transforming to the continuous limit we get the
differential equation
\begin{equation}
{\partial \theta^2\over \partial\phi}=-\,{2 \over \psi
(\theta)}\,\,{R_c+a\over R_ca}\,\,{\bbox\delta}(\phi){\bf e}(\phi)~,
\label{e49}
\end{equation}
whose solution, after substituting $\psi(\theta)$ from Eq.~(\ref{e47}),
yields
\begin{equation}
\theta^3(\phi)=\theta^3(\phi_i)-{3\over 2a}\left({R_c+a\over
R_c}\right)^2\int_{\phi_i}^{\phi}\!\!d\phi'
{\bbox\delta}(\phi'){\bf e}(\phi')~.
\label{e50}
\end{equation}
The function ${\bbox\delta}(\phi)$ in Eqs.~(\ref{e49}),(\ref{e50}) gives
the drift shift for the cyclotron orbit whose guiding center is a
distance $R_c+a$ from the center of the disc in the direction
specified by the angle $\phi$. In fact, the integration in
Eq.~(\ref{e50}) runs along the guiding-center trajectory during the
rollover, which is the contour $\rho=R_c+a$, where $\bbox\rho$ is the
radius vector counted from the center of the disc. Note that the
integral term in Eq.~(\ref{e50}) is bounded from above by $\sim
\delta/a$ at $R_c\gg a$ and by $\delta a/R_c^2\sim a/(dl_L)^{1/2}$
otherwise. In both limits the maximum value of $\theta\ll 1$, which
justifies the linearization (\ref{e47}) and our using of the term
``rollover".

If we take $\phi_i$ in Eq.~(\ref{e50}) equal to the polar angle
at which the cyclotron orbit hits the disc for the first time, then
$\theta(\phi_i)$ should be put to zero in the continuous
approximation. With the same accuracy the angle $\phi_f$ at which the
breakaway occurs satisfies the condition $\theta (\phi_f)=0$.
According to Eq.~(\ref{e50}), we can recast the latter condition as
\begin{equation}
\int_{\phi_i}^{\phi_f} \!\!d\phi \,\,{\bbox\delta}(\phi){\bf
e}(\phi)=0~.
\label{e51}
\end{equation}  
This equation yields $\phi_f$ as a function of $\phi_i$ and,
consequently, enables us to determine the shift $R_h$. Let us
introduce an effective random potential $V({\bbox\rho})$ as the
average of the real potential over the cyclotron orbit with the guiding
center at the point $\bbox\rho$ [at $R_c\ll d$ the two potentials
almost coincide, but at $R_c\gg d$ a typical amplitude of fluctuations
of the effective potential $V({\bbox\rho})$ with the same correlation
radius $d$ is obviously $\sim (R_c/d)^{1/2}$ times smaller]. The drift
occurs along equipotential lines of $V({\bbox\rho})$. Since the
shift ${\bbox{\delta}}({\bbox\rho})\propto \nabla V({\bbox\rho})\times
{\bf e}_z$, where ${\bf e}_z$ is the unit vector along the magnetic
field, Eq.~(\ref{e51}) is rewritten as
\begin{equation}
\int_{\phi_i}^{\phi_f}\!\!d\phi\,\,(\nabla V\times {\bf e}_z)\cdot
{\bf e(\phi)}=0~,
\label{e52}
\end{equation}
which, for the integration along the arc $\rho=R_c+a$, finally gives
\begin{equation} 
V(\rho,\phi_f)-V(\rho,\phi_i)=0~.  
\label{e53}
\end{equation}
We thus see that in the limit of small $\delta/a$ the cyclotron orbit,
having skipped along the surface of the hard disc, breaks away on the
equipotential contour with the same $V$ as it had before hitting the disc,
i.e., 
\begin{equation}
R_h=0
\label{e54}
\end{equation}
in the continuous approximation. Note that for the drift in a
homogeneous electric field, when ${\bbox\delta}({\bbox\rho})={\rm
const}$, this result follows straightforwardly from symmetry of the
scattering problem. What Eq.~(\ref{e53}) tells us is that, remarkably,
$R_h$ vanishes in the case of varying $V({\bbox\rho})$ as well.

Let us compare the above picture with a familiar example of
adiabaticity of scattering on a smooth inhomogeneity $V({\bbox\rho})$:
in that case, the vanishing of $R_h$ simply means that the guiding
center drifts along a locally perturbed equipotential line of
$V({\bbox\rho})$. Naively, one might think that a collision with the
disc destroys the adiabaticity since the impenetrable hard disc makes
a part of the equipotential line of $V({\bbox\rho})$ inaccessible.
However, as follows from Eq.~(\ref{e53}), the skipping of the cyclotron
orbit around the disc goes on adiabatically, provided $\delta/a$ is
infinitesimally small, and the result $R_h=0$ still holds.

\subsection{Nonadiabatic skipping}
\label{nonadskip}

The zero result for $R_h$ was obtained in Sec.~\ref{skipor} by
treating the drift during the rollover perturbatively, to first order
in $\delta/a$, and by taking the continuous limit. To find $R_h$, we
should now relax this approximation. Before doing so, it is worthwhile
to recall how $R_h$ behaves for scattering on a smooth
inhomogeneity. In that problem, it is known that taking higher
gradient terms into account does not lead to a finite $R_h$ and, in
fact, the problem of finding $R_h$ does not allow for any perturbative
solution that could be expanded in powers of the parameter
$\delta/d\ll 1$, where $d$ is a characteristic size of the
inhomogeneity. Specifically, for a smooth inhomogeneity, $R_h$ is
exponentially small at $\delta/d\ll 1$ (see, e.g., \cite{evers99a} for
a solution of the scattering problem and references therein). As we
have shown above, the case of a hard disc placed in a smoothly varying
environment is similar in that the scattering is also almost adiabatic
at $\delta/a\ll 1$. A question then arises if the nonadiabatic
scattering that leads to a finite $R_h$ is also exponentially
suppressed. The answer is no, since there is an important difference
between the two cases. Namely, the approximation within which the
skipping of the cyclotron orbit can be considered as a continuous
adiabatic process [Eq.~(\ref{e49})] fails completely near the points
$\phi=\phi_i$ and $\phi=\phi_f$. Indeed, near these points dynamics of
the collision angle $\theta (\phi)$ is nonadiabatic since $\theta$ is
close to zero, so that $\Delta\theta_n$ is of order $\theta_n$ itself
and the expansion (\ref{e48}) is not valid any more. A finite shift
$R_h$ is therefore due to the discreteness and incommensurability of
the skipping along the sharp boundary of the hard disc. It is given by
the elementary (associated with a single collision) hopping length of
the guiding center near $\phi=\phi_{i,f}$, which is $R_{h1}\sim
(R_c+a)\psi (\theta_1)$. Substituting $\theta_1$ from Eq.~(\ref{e44})
we finally get for a characteristic amplitude of the shift
\begin{equation}
R_{h1}\sim R_c\left({\delta\over a}\,{R_c+a\over R_c}\right)^{1/2}~.
\label{e55}
\end{equation}
We see that, in contrast to the case of a smooth inhomogeneity,
$R_{h1}$ scales as a power of $\delta$, namely $R_{h1}\propto
\delta^{1/2}$. Note also that the characteristic scale $d$, on which
the smooth $V({\bbox\rho})$ changes, appears in Eq.~(\ref{e55}) only
through the drift shift $\delta$.

Equation (\ref{e55}) describes a single scattering in which the
particle breaks away from the disc along an equipotential with $V$
almost equal to that of the equipotential along which it is incident
on the disc. Using Eq.~(\ref{e55}), dynamics of skipping orbits in the
limit $R_c\ll d$ can be understood quite straightforwardly. Since in
this limit the drift is almost homogeneous in the course of skipping,
for a given $V$ there is typically only one equipotential line that
crosses the guiding-center trajectory during the
rollover. Accordingly, if it were not for the small shift (\ref{e55}),
the particle would simply continue to drift after the rollover along
the same equipotential line. The picture becomes far more complicated
at large $R_c\gg d$ since in that case there are many equipotential
lines with the same $V$ that intersect the guiding-center trajectory
corresponding to the rollover.

\begin{figure}
\begin{center}
\includegraphics[width=0.9\columnwidth,clip]{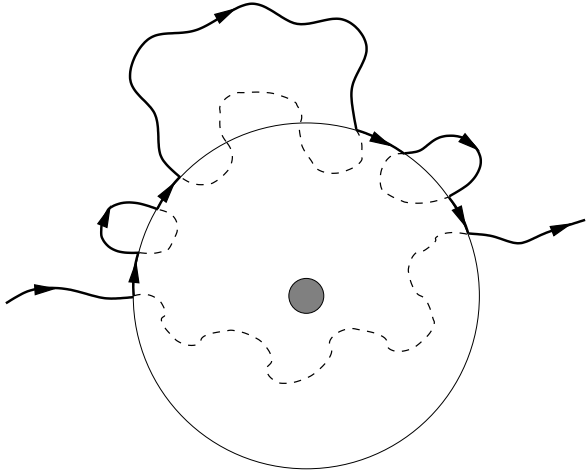}
\end{center}
\vspace{3mm}
\caption{Schematic picture of scattering of a cyclotron orbit on an
antidot at $\delta\ll a$, $R_c\gg d$ and $R_{h\Sigma}\ll w_S$. The
thick solid line with arrows is the guiding-center trajectory. The
dashed lines are equipotential contours which cannot be accessed by
the guiding center. The small shaded circle in the center of the
figure shows the antidot. The thin circle of radius $R_c+a$ around it
is the boundary of the area impenetrable for the guiding center. The
parts of the guiding-center trajectory that coincide with this
boundary correspond to skipping of the cyclotron orbit, which
alternate with parts corresponding to drift. The drift occurs between
each breakaway from the antidot and consecutive return to it. }
\label{break}
\end{figure}

Let $R_c\gg d$. In this limit, one should distinguish two regimes
according to whether the typical hopping length after one collision
$R_{h1}\sim R_c (\delta/a)^{1/2}$ [Eq.~(\ref{e55})] is larger or
smaller than $d$. Consider first the case $R_{h1}\ll d$, i.e., let
$d\ll R_c\ll d(a/\delta)^{1/2}$. The characteristic number of
equipotential lines with the same $V$ that cross the circle of radius
$\rho=R_c+a$ around the disc (which is the guiding center trajectory
in the course of skipping) is of order $R_c/d\gg 1$. Clearly, the
direction of drift (to or away from the surface of the disc)
alternates during the skipping. It follows that the particle which
started skipping will break away along the equipotential line that is
the first to cross the guiding-center trajectory corresponding to the
skipping. Yet, since the equipotential lines are closed loops
(typically of size $d$), the particle will come full circle and return
to the disc (see Fig.~\ref{break}). Then the process will repeat
itself with other equipotential lines along the surface of the
disc. We have assumed, however, that the particle is incident on the
disc along a percolating extended trajectory. Therefore, the multiple
returns will stop when the particle picks up this trajectory.

Let us evaluate the total shift $R_{h\Sigma}$ with which the particle
will finally break away. Elementary shifts for the repeating
collisions are uncorrelated with each other, so that
\begin{equation}
R^2_{h\Sigma}\sim R^2_{h1}R_c\left[\int_d^{R_c}\!\!d\Lambda\,
\Lambda\, W(\Lambda)\right]^{-1}~,
\label{e56}
\end{equation}
where $W(\Lambda)$ is the probability density for the drift trajectory
that broke away from the disc to hit it again, i.e., to cross the arc
$\rho=R_c+a$, for the first time after the breakaway at a distance
$\Lambda$ from the starting point. It is instructive to map the
problem of finding $W(\Lambda)$ onto a more conventional one by noting
that the power-law scaling of $W(\Lambda)$ describes how a deposition
rate for particles emitted by a point source and moving in two
dimensions in the presence of an absorbing line falls off with
increasing distance $\Lambda$ along this line. If the particles would
experience an uncorrelated diffusion with the elementary step $\sim
d$, then it is straightforward to see, by solving the diffusion
equation with the absorbing boundary, that $W(\Lambda)\sim
d/\Lambda^2$. In fact, one can show, by introducing a scale-dependent
diffusion coefficient which describes the drift, that this result
holds for the drifting particles as well, i.e., $W(\Lambda)$ and
$P(\xi)$ [Eq.~(\ref{e19})] have the same scaling behavior. Hence, the
integral in Eq.~(\ref{e56}) logarithmically diverges and the typical
number of collisions before the final breakaway is $R_c/d\ln
(R_c/d)$. It follows that
\begin{equation} 
R_{h\Sigma}\sim R_c\left[{\delta\over a}\,{R_c\over
d\ln (R_c/d)}\right]^{1/2}~.  
\label{e57} 
\end{equation} 
Clearly, the necessary condition for the above derivation of
$R_{h\Sigma}$ to be valid is $R_{h\Sigma}\ll d\ll R_c$, which means
$d\ll R_c\ll d(a/\delta)^{1/3}\ln^{1/3}(a/\delta)$. It is worthwhile
to mention that if $R_c\ll d(a/\delta)^{1/3}$ the
characteristic hopping length for the skipping cyclotron orbit
$R_c\psi_{max}\ll d$. Here $\psi_{max}\sim (\delta/a)^{1/3}$ is the
maximum scattering angle for a single-run skipping at $R_c\gg a$ [see
Eqs.~(\ref{e47}),(\ref{e50})].

Now let $R_{h1}$ be still smaller than $d$ but let $R_{h\Sigma}\gg d$,
i.e., consider the interval $d(a/\delta)^{1/3}\ln^{1/3}(a/\delta)\ll
R_c\ll d(a/\delta)^{1/2}$. An essential difference appears in this
regime: the shift accumulated through multiple breakaways and returns
exceeds $d$ before the rollover is finished. At this point it is worth
recalling that the elementary shifts for each breakaway are
accompanied by changes of $V$ corresponding to drift trajectories.
Accordingly, when the accumulated shift gets larger than $d$, the
initial and current values of $V$ become uncorrelated with each other.
It is evident that we should treat this case separately, since now one
cannot identify $R_{h\Sigma}$ given by Eq.~(\ref{e57}) with a shift
with which the particle has finally broken away never to return. This
conclusion becomes even clearer when not only $R_{h\Sigma}$ but also
$R_{h1}$ [Eq.~(\ref{e55})] is larger than $d$, i.e., when $R_c\gg
d(a/\delta)^{1/2}$. In the latter case, the memory about the initial
value of $V$ is lost already after one collision.

\begin{figure}
\begin{center}
\includegraphics[width=0.9\columnwidth,clip]{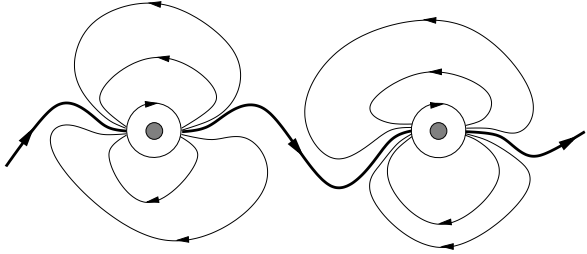}
\end{center}
\caption{Schematic picture of scattering of a cyclotron orbit on
antidots at $\delta\ll a$, $R_c\ll d$ and $R_{h1}\gg w_S$. The line
with arrows shows the guiding-center trajectory. The thicker line is
the trajectory that takes the particle from one antidot to another
(two of which are shown by shaded circles). Until the guiding center
picks up this trajectory the particle is stuck to an antidot and keeps
colliding with it, by making long drift excursions between the
collisions. For $R_c\gg d$, the scattering process can be visualized
as a ``combination" of Figs.~\ref{break},\ref{returns}. }
\label{returns}
\end{figure}

The picture emerging in the limit $R_{h\Sigma}\gg d$ signals that one
should also fine-tune the derivation of
Eqs.~(\ref{e55}),(\ref{e57}). Namely, these equations should be
supplemented with the condition under which the problem of finding
$R_h$ can in fact be formulated as the scattering problem for a {\it
single} disc. Indeed, for a single disc, one actually cannot specify
with which accuracy the particle should hit the vicinity of the
percolating trajectory so as to be able to break away from the
disc. To answer this question, we need to consider a scattering
problem for {\it two} discs. Specifically, consider two discs
separated by a typical distance $L_S\sim v_d\tau'_S$ measured along
the percolating trajectory (Fig.~\ref{returns}). The two discs are
connected by a bundle of drift trajectories of width $w_S$
[Eq.~(\ref{e31})].  One sees that the particle will break away from
one disc and get through to the other after a single rollover only if
$\max \{R_{h1},R_{h\Sigma}\}$ is within this width. If this condition
is not satisfied, the cyclotron orbit will keep going round the disc
until it happens that its guiding center hits the strip of width $w_S$
at the same time when $\theta$ vanishes. It follows that at $R_c\ll d$
the shift is given by Eq.~(\ref{e55}) only if $R_{h1}\ll w_S$. If,
however, $R_{h1}\gg w_S$, it takes typically $N_r\sim R_{h1}/w_S$
revolutions around the disc before the particle breaks away and
\begin{equation}
R_h\sim w_S
\label{e58}
\end{equation} 
in this limit. Similarly, at $R_c\gg d$ the shift $R_h$ is given by
Eq.~(\ref{e57}) only if $R_{h\Sigma}\ll w_S$; otherwise, $R_h\sim
w_S$ (the picture can be visualized as a ``combination" of
Figs.~\ref{break},\ref{returns}). For $R_c\gg d$, the number of
revolutions around the disc $N_r$, necessary for the simultaneous
tuning of $V$ and $\theta$ corresponding to the breakaway, is
different in three different regimes. Namely, $N_r\sim
R_{h\Sigma}/w_S$ for $w_S\ll R_{h\Sigma}\ll d$, independently of the
ratio $R_{h1}/w_S$. With increasing $R_{h\Sigma}$, if $R_{h\Sigma}\gg
d$, we have $N_r\sim d/w_S$ as long as $R_{h1}\ll d$, and $N_r\sim
R_{h1}/w_S$ otherwise. In particular, one sees that if $R_{h\Sigma}\gg
d$, it always takes many revolutions before the breakaway occurs and
the shift $R_h$ is always determined by $w_S$.

\subsection{Percolation in a dilute antidot array}
\label{rare}

In Sec.~\ref{nonadskip}, we discussed scattering of a drifting
cyclotron orbit on a single AD at $\delta/a\ll 1$. Consider now the MR
at $\delta/a\ll 1$. Let us start with the limit of a small
concentration of ADs $n\to 0$. As outlined in Sec.~\ref{outline2}, in
the absence of ADs the conductivity at large $B$ is only due to the
exponentially weak nonadiabatic scattering on long-range disorder. Let
us neglect this additional contribution to $\rho_{xx}(B)$ and
calculate the contribution that is due to the scattering on very rare
hard discs. We thus seek a term in $\rho_{xx}(B)$ which is
proportional to a power of $n$ at small $n$. Let $R_c\ll d$. Since at
$n\to 0$ the bundle of drift trajectories that connect two discs
becomes infinitesimally narrow [$w_S\propto n^{3/7}$ at $n\to 0$
according to Eqs.~(\ref{e27}),(\ref{e31})], the particle sticks to a
disc for a long time $\tau_{st}$ until it picks up a trajectory that
is extended enough to take it to another disc. The number of the
unsuccessful attempts to break away from the disc is given by $N_r\sim
R_{h1}/w_S\gg 1$. To evaluate $\tau_{st}$, notice that the sticking
time is determined by a slow drift along the closed loops that
repeatedly return the particle back to the disc, not by the fast
skipping in between. However, most of the attempts end up in quick
returns to the vicinity of the point of the first collision and so
give only a small contribution to $\tau_{st}$. We estimate $\tau_{st}$
as
\begin{equation}
\tau_{st}\sim\int_{w_S}^{R_{h1}}\!{dw\over w_S}\,{L(w)\over v_d}~,
\label{e59}
\end{equation} 
where $L(w)$ is given by Eq.~(\ref{e7}). The integral is determined by
$w\sim w_S$, which yields $\tau_{st}\sim \tau'_S$. We thus see that
the effective scattering time between collisions with different discs
is given by the drift time between the discs. Accordingly, the total
distance passed by the particle in the multiple rollovers is typically
of order $L_S$. In fact, this conclusion holds for the case $R_c\gg d$
as well. The only difference is that at large $R_c$ the particle
experiences multiple breakaways and returns on the scale of a single
rollover. However, $\tau_{st}$ does not renormalize the characteristic
time between collisions with different discs, which is given by
$\tau'_S$, similarly to the case $R_c\ll d$.

We are now in a position to calculate the AD-induced contribution to
the MR in the limit of small $n$. According to the picture above, the
scattering on a given AD is over when the particle hits another AD
separated by a distance $\xi_S\sim d(v_d\tau'_S/d)^{4/7}$, and the
characteristic time between two scatterings is $\tau'_S$. As long as
we neglect the nonadiabatic corrections to the drift in the long-range
potential, there is a clear separation between localized and extended
drift trajectories. Only a small fraction of trajectories get
delocalized by means of collisions with ADs, namely $\sim
w_SL_S/\xi_S^2\ll 1$. The diffusion coefficient of particles residing
on these trajectories is $\sim v_d\xi_S^2/L_S$. We thus see that the
macroscopic diffusion coefficient is given by Eq.~(\ref{e9}) with a
rescaled scattering time $\tau_S\to \tau'_S$, which yields the MR
obeying Eq.~(\ref{e42}). 

It is worth noting that the dynamics of particles is essentially
different at $n\to\infty$ (Sec.~\ref{hydro}) and $n\to 0$, despite
being described by similar equations. In the former case, there is a
strong exchange, caused by collisions with ADs and governed by a
detailed balance of scattering processes, between the stream of fast
particles which follow the links of the percolation network and the
reservoir of ``quasilocalized" particles which stick for a long time
to within the critical cells of the network. By contrast, at $n\to 0$
there is no such exchange and the drift trajectories within the cells
of the percolation network are strictly localized. However, in both
cases the MR is determined by the fast particles moving along the
links of the percolation network, which is why it does not matter if
the particles moving inside the critical cells are localized or
not. Put another way, although the total number of delocalized
particles decreases at $n\to 0$, this effect is compensated by more
frequent crossings, due to collisions with ADs, of the percolative
drift trajectory by particles that remain delocalized.

Substituting Eqs.~(\ref{e27}),(\ref{e33}) into
Eq.~(\ref{e42}) we get
\begin{equation}
{\rho_{xx}(B)\over\rho_0}\sim p^{4/7}\eta^{3/7}\,\left({d\over
R_c}\right)^{39/28}
\label{e60}
\end{equation}
for $R_c\gg d$ and 
\begin{equation}
{\rho_{xx}(B)\over\rho_0}\sim p^{4/7}\eta^{3/7}\,\left({d\over
R_c}\right)^{4/7}
\label{e61}
\end{equation}
in the opposite limit. According to Eqs.~(\ref{e60}),(\ref{e61}),
$\rho_{xx}\propto n^{3/7}$ at $n\to 0$. 

One sees from Eq.~(\ref{e61}) that $\rho_{xx}$ taken in the limit $n\to
0$ behaves at large $B$ as $B^{4/7}$. As compared to Eq.~(\ref{e17}),
which describes the asymptotic behavior in the hydrodynamic regime
$n\to \infty$ and gives $\rho_{xx}\propto B^{10/13}$, the divergence
of $\rho_{xx}$ with increasing $B$ is weakened, but is still
present. However, neither of the two limiting cases
(\ref{e17}),(\ref{e61}) describes the asymptotics of the MR for
$B\to\infty$ at a given finite $n$, which we discuss below.

Let us turn to the percolative MR in a denser array of ADs.
Increasing $n$ yields wider links of the percolation network, so that
eventually $w_S$ becomes larger than the elementary shift
$\max\{R_{h1},R_{h\Sigma}\}$ (Sec.~\ref{nonadskip}). This transport
regime is described by the advection-diffusion equation (\ref{e17})
modified according to Eq.~(\ref{e43}). Using Eqs.~(\ref{e55}),(\ref{e57})
for $R_{h1}$ and $R_{h\Sigma}$, and Eqs.~(\ref{e27}),(\ref{e33}) for
$\tau'_S$ in Eq.~(\ref{e43}) gives
\begin{equation}
{\rho_{xx}(B)\over\rho_0}\sim p^{10/13}\eta^{6/13}\left({d\over
R_c}\right)^{1/52}\ln^{-3/13}\left({R_c\over d}\right)
\label{e62}
\end{equation}
for $R_c\gg d$ and
\begin{equation}
{\rho_{xx}(B)\over\rho_0}\sim p^{10/13}\eta^{6/13}\left({R_c\over
d}\right)^{2/13}\left({R_c+a\over
R_c}\right)^{3/13}
\label{e63}
\end{equation}
for $R_c\ll d$. 

The condition at which the ``one-hop" percolation
[Eqs.~(\ref{e60}),(\ref{e61})] crosses over with increasing $B$ into the
advection-diffusion regime [Eqs.~(\ref{e62}),(\ref{e63})] is given by
somewhat cumbersome formulas: the crossover occurs at $d/R_c\sim
(p^6\eta)^{3/125}\ln^{-21/125}(p^6\eta)$ for $p^6\eta\ll 1$, at
$d/R_c\sim (p^6\eta)^{1/22}$ for $1\ll (p^6\eta)^{1/22}\ll d/a$, and
at $d/R_c\sim (p^6\eta)^{1/15}(a/d)^{7/15}$ for $(p^6\eta)^{1/22}\gg
d/a$.

Equation (\ref{e63}) tells us that the asymptotics of the MR at
$B\to\infty$ is $\rho_{xx}\propto B^{1/13}$, i.e., the MR diverges as
a power law in the limit of large $B$. However, this divergence is so
weak that from a practical point of view it is indistinguishable
from a saturation of $\rho_{xx}(B)$. Nonetheless, it is a remarkable
fact that in the extreme $B\to\infty$ the MR in the presence of both
ADs and a long-range potential does not go to zero, in contrast to
Eqs.~(\ref{e1}),(\ref{e4}), which predict vanishing of $\rho_{xx}$ when
only one type of disorder is present.

\subsection{Rosette orbits in the presence of weak drift}
\label{short}

In Sec.~\ref{rare}, we considered the percolative contribution to the
MR at small $\delta/a$. Now we proceed to the ``short-scale"
contribution associated with rosette states
(Secs.~\ref{outline1},\ref{finn}). Let us analyze how the Lorentz-gas
behavior is restored with decreasing strength of the smooth
disorder. Since we deal with the case $a/d\ll 1$, the Lorentz-model
limit is achieved for a very weak long-range potential, such that at
$R_c/l_S\sim 1$, when the falloff (\ref{e1}) starts, the scattering on
the long-range potential is already strongly adiabatic. The
condition of adiabaticity at $R_c/\l_S\sim 1$ and $l_S/d\gg 1$
translates into $l_L/d\gg (l_S/d)^3$. As we will see below, the
condition of the Lorentz-gas falloff $\rho_{x}(B)\propto B^{-1}$ not
being dominated by the contribution of ``hopping rings"
(Sec.~\ref{finn}) is much stronger, namely
\begin{equation}
l_L/d\gg (l_S/d)^3(d/a)^4~.
\label{e64}
\end{equation}

The solution of the scattering problem for a single disc in
Sec.~\ref{skipor} shows that at $\delta/a\ll 1$ there is a
well-defined separatrix $\theta_{max}\ll 1$ for the angle of incidence
$\theta$ in the phase space $(\theta,\phi)$, which divides skipping
cyclotron orbits into two groups, delocalized and localized. Namely,
trajectories that will finally break away from the disc belong to the
part of the phase space with $\theta<\theta_{max}$, whereas the region
$\theta>\theta_{max}$ is filled with those that will never escape. In
other words, almost the whole phase space is filled with bound
states. According to Eq.~(\ref{e50}), the critical angle
\begin{equation} 
\theta_{max}\sim
(\delta/a)^{1/3}[(R_c+a)/R_c]^{2/3}~.  
\label{e65} 
\end{equation}
Indeed, a drifting cyclotron orbit that is incident on the disc both
hits it for the first time and finally breaks away at
$\theta\alt\theta_1$ [Eq.~(\ref{e44})], while reaching a maximum
$\theta$, which depends on the incident parameter but does not exceed
$\theta_{max}$ given by Eq.~(\ref{e65}), during a rollover in
between. 

Let us show that a particle which starts with some
$\theta>\theta_{max}$ will not be able to break away. What is
important to us is that in order to break away the particle has to
decrease $\theta$ down to $\theta\sim \theta_1$. Suppose this might
happen and the particle has escaped. Then we could consider a
complementary scattering problem by reversing time and sending the
particle that has broken away back to the disc. However, we know from
Sec.~\ref{skipor} that for this scattering problem $\theta$ will never
exceed $\theta_{max}$. Hence, we come to a contradiction which shows
that the initial assumption about the possibility of a breakaway
cannot be realized. We thus see that trajectories that started to skip
along the disc with $\theta>\theta_{max}$ can never cross the boundary
$\theta=\theta_{max}$ and so will remain bound to the disc. The fact
that there exist bound states for magnetized electrons interacting
with a single hard disc in the case of a homogeneous in-plane electric
field was observed in the numerical simulation \cite{berglund96}.

The existence of the separatrix $\theta_{max}$ means that the
contribution to $\rho_{xx}(B)$ of rosette states with
$\theta>\theta_{max}$ is given by 
\begin{equation}
\rho_{xx}(B)/\rho_0\sim R_c/l_S 
\label{e66} 
\end{equation} 
for all $\delta/a\ll 1$ [cf.\ Eq.~(\ref{e1})].  

Let us now compare two terms in $\rho_{xx}(B)$ that are associated
with the short-scale diffusion at finite $\delta/a$. One, described by
Eq.~(\ref{e22}), is due to the hopping of ``rings" introduced in
Sec.~\ref{finn}. The other, given by Eq.~(\ref{e66}), is due to the
hopping of rosette states. Note a similarity between the two
mechanisms of transport: in both cases a particle interacts with a
disc many times before changing to another disc. The comparison shows
that the concentration of particles participating in the hopping-ring
transport is much larger than that of rosette states. It is clear,
however, that if we send $\delta\to 0$, the hopping rings should not
contribute to $\rho_{xx}(B)$, which will be given by Eq.~(\ref{e1}). It
follows that there should exist yet another, intermediate regime of
hopping, associated with the evolution of the hopping-ring transport
with decreasing $\delta/a$. To describe the latter, notice first of
all that Eq.~(\ref{e22}) stops to be valid already at some $\delta/a\gg
1$. Indeed, the logarithmic factor in Eq.~(\ref{e22}) is associated
with the drift along closed loops of size $\xi\alt
(nR_c)^{-1}$. Therefore, the derivation of Eq.~(\ref{e22}) in fact
implies that the time it takes to come full circle along the longest
loop of size $\xi\sim (nR_c)^{-1}$ is smaller than $\tilde\tau_S(\xi)$
[Eq.~(\ref{e21})] for this $\xi$, which yields the condition
$\delta/a\agt (nR_cd)^{-3/4}\gg 1$. Note that this is the same
condition at which the effective scattering time for $d\ll \xi_S\ll
R_c$ in Sec.~\ref{scatt} is not renormalized by the parameter
$\delta/a$ and is given by $\tau_S$ [cf.\ Eq.~(\ref{e30})]. At smaller
$\delta/a$, in the interval $1\ll \delta/a\ll (nR_cd)^{-3/4}$, only
$\xi$ in the range $1\alt \xi/d\alt (\delta/a)^{4/3}$ contribute to
$\rho_{xx}$, which gives 
\begin{equation} 
\rho_{xx}(B)/\rho_0\sim nR_cd\,\ln (\delta/a)~.  
\label{e67} 
\end{equation} 
This equation is valid with decreasing $\delta/a$ down to
$\delta/a\sim 1$. At still smaller $\delta/a$, the particle is
scattered out by the same AD after passing a distance of order
$d$. Accordingly, the characteristic time between changes of ADs in
this regime increases due to the slowing down of the drift as
$\tau_Sa/\delta$. It follows that for $\delta/a\ll 1$ 
\begin{equation}
\rho_{xx}(B)/\rho_0\sim nR_cd\,\delta/a~.  
\label{e68} 
\end{equation}

Equations (\ref{e68}),(\ref{e66}) match each other at $\delta\sim a^2/d$
and we conclude that the regimes (\ref{e67}),(\ref{e68}), intermediate
between those described by Eqs.~(\ref{e22}) and (\ref{e66}), occur in
the interval $a/d\ll\delta/a\ll (nR_cd)^{-3/4}$. By requiring that
$\delta$ at $R_c/l_S\sim 1$ is much smaller than $a^2/d$ we arrive at
the condition (\ref{e64}). The overall behavior of $\rho_{xx}(B)$ for
the case of very weak long-range disorder is obtained by adding the
``short-scale" contribution, described by
Eqs.~(\ref{e22}),(\ref{e66})-(\ref{e68}), and the percolative
contribution analyzed in Sec.~\ref{rare}. This leads to nonmonotonic
behavior of the MR, such that $\rho_{xx}(B)$ first falls off with
increasing $B$ and then crosses over into the percolative growth.

\section{Numerical simulation}
\label{numer}

We have solved numerically the classical equation of motion for a
charged particle in a random array of hard discs in the presence of
smooth disorder. In the numerical simulation, the latter is
characterized by the correlator
\begin{equation}
\left<V({\bf r})V(0)\right>\propto {1\over [1+(r/d)^2]^{3/2}}
\label{e69}
\end{equation}
with the ratio $d/a\simeq 21.5$. The concentration $n$ of the isolated
scatterers has been chosen such that they make an important
contribution to the resistance at zero magnetic field. Specifically,
$\pi nd^2\simeq 6.2$ and $l/l_S\simeq 0.58$, where $l$ is the total mean
free path and $l_S=3/8na$.

\begin{figure}
\begin{center}
\includegraphics[width=0.95\columnwidth,clip]{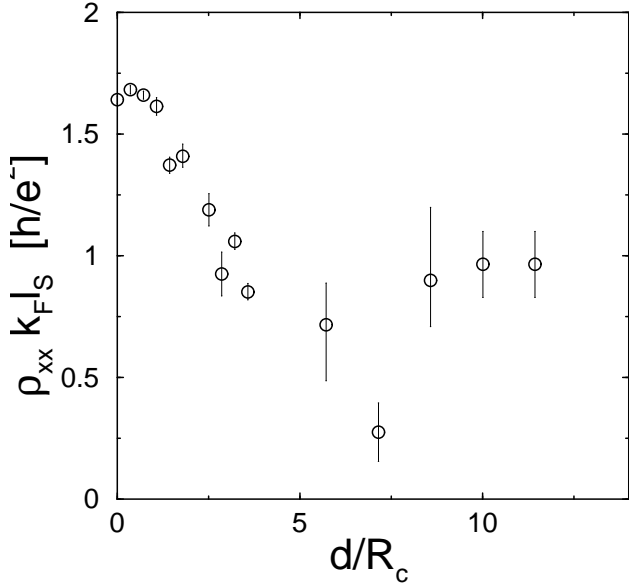}
\end{center}
\caption{Magnetoresistivity $\rho_{xx}(B)$ in units of
$(h/e^2)/k_Fl_S$ (spin included) as a function of $d/R_c$ for a model
system specified in the text. Characteristic values of $d/R_c$ are:
$R_c/l_S=1$ at $d/R_c\simeq 0.25$; $\delta/d=1$ at $d/R_c\simeq 1.8$;
$nR_cd=1$ at $d/R_c\simeq 1.9$; $\pi nR_c^2=1$ at $d/R_c\simeq 2.5$;
$\delta/2a=1$ at $d/R_c\simeq 7.1$.}
\label{numerfig}
\end{figure}

In Fig.~\ref{numerfig} we present the MR data for our model system.
Since the characteristic values of the magnetic field (given in the
figure caption in terms of the parameter $d/R_c$) are rather close to
each other, it has not been possible to unambiguously separate
different regimes. Yet, the nonmonotonic behavior of $\rho_{xx}(B)$,
predicted by the theoretical analysis, is clearly seen. Note that the
size of the error bars at $d/R_c\agt 5$ denotes only the statistical
uncertainty.  In addition to the latter, there is a systematic
uncertainty originating from a very slow guiding-center motion in the
large-$B$ limit, which makes it difficult to observe the true
diffusion constant. However, the size of the systematic uncertainty is
sufficiently small as compared to the structure of the nonmonotonic
dependence of $\rho_{xx}(B)$.

\section{Summary}
\label{conclu}

In summary, we have discussed a rich set of magnetotransport phenomena
which take place in a random ensemble of antidots in the presence of
long-range fluctuations of a random potential. We believe that the
model studied in the paper adequately describes an antidot array in
semiconductor heterostructures with a wide spacer for not too high
$B$, when quantum effects (Shubnikov-de Haas oscillations) are still
weak. We show that even weak long-range disorder yields a wealth of
pronounced effects in the behavior of the magnetoresistance
$\rho_{xx}(B)$ in the antidot array. Essentially, these effects are
associated with a magnetic-field induced localization of electrons
which develops when only one type of disorder is present. As a result
of the localization, $\rho_{xx}(B)$ vanishes in the limit of large $B$
both in an idealized antidot system without long-range disorder and in
a system with smooth inhomogeneities without antidots. Conceptually,
the most striking result of our work is that the interplay of two
types of disorder does not simply modify the localization; in fact, it
destroys the localization, so that $\rho_{xx}(B)$ even diverges in the
limit $B\to \infty$. The divergence takes place despite a strong
falloff of $\rho_{xx}(B)$ that occurs in intermediate magnetic fields
in the case when one of the types of disorder is sufficiently strong
as compared to the other.

Piecing together all the numerous regimes we arrive at a rather
complex overall picture, due to the interplay of several distinctly
different mechanisms of the MR. Let us list these mechanisms. Some of
them are closely related to the mechanisms of the MR characteristic to
the limiting cases of strongly non-Gaussian or purely Gaussian
disorder. Specifically, we have:

(a) Memory effects operative in the case of strongly non-Gaussian
disorder (Lorentz gas, or any other system of rare strong scatterers,
without long-range inhomogeneities). These effects lead to a strong
{\it negative} MR (see Sec.~\ref{outline1}). In the limit of large $B$
the system is insulating.

(b) Memory effects in smooth Gaussian disorder. In this case, the
memory effects give rise to a strong {\it positive} MR, for which,
however, the ratio $\rho_{xx}(B)/\rho_0$ cannot be parametrically much
larger than 1, since after having reached a maximum value of order 1
it starts to fall off with increasing $B$, thus yielding a strong {\it
negative} MR (see Sec.~\ref{outline2}). In the limit of large $B$ the
system is insulating.

In the presence of a long-range random potential, the mechanism (a) of
the MR is destroyed by the diffusive motion of electrons scattered by
the long-range disorder. Nonetheless, as we have demonstrated in the
paper, it is reincarnated in the form of ``hopping rings"
(Sec.~\ref{finn}) once the diffusive motion in the long-range disorder
turns into the drift with increasing $B$. In this new form, this kind
of a negative contribution to the MR is developed at much larger $B$
as compared to the pure Lorentz-gas system. On the other hand, the
mechanism (b) is destroyed by scattering on hard scatterers, which
checks the negative MR associated with the adiabaticity of drift in
the long-range potential (Sec.~\ref{nonad}).

A nontrivial point to notice in our results is that, in addition to
the above, there are memory effects that are specific to the
inhomogeneous system with two types of disorder. These are:
 
(c) ``Diffusion-controlled percolation". As we have shown in the
paper, scattering by short-range inhomogeneities not just destroys the
adiabaticity of motion in a smooth random potential, thus checking the
strong negative MR. In fact, it reverses the sign of the MR by giving
rise to a {\it positive} MR which keeps growing with increasing $B$
(Sec.~\ref{hydro}). This positive MR is a peculiar feature of
percolation of drift trajectories with superimposed diffusive dynamics
across the drift lines.

(d) Renormalization, by long-range disorder, of the collision time
for hard scatterers. We have demonstrated that in a system where the
hard scatterers give the main contribution to the scattering rate at
zero $B$, a weak smooth disorder can drastically suppress the
scattering rate with increasing $B$ (Sec.~\ref{scatt}). The effect
takes place for any type of dynamics of scattering by the long-range
potential, both for diffusion and drift. The increase of the collision
time translates into the {\it negative} MR.

\begin{figure}
\begin{center}
\includegraphics[width=0.95\columnwidth,clip]{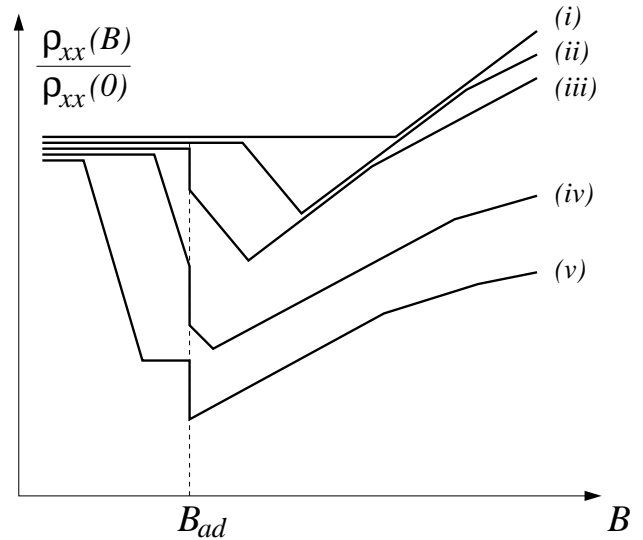}
\end{center}
\vspace{3mm}
\caption{Schematic behavior of the magnetoresistivity $\rho_{xx}(B)$
on a log-log scale for different values of the concentration of
antidots $n$: $n^{(i)}>n^{(ii)}>\ldots >n^{(v)}$, keeping all other
parameters ($l_S,l_L,d$) fixed. Only one characteristic field $B_{ad}$
is shown, at which the crossover between diffusive dynamics and
adiabatic drift in the long-range potential takes place. Different
curves illustrate different mechanisms of the magnetoresistance: (i)
the magnetoresistance is positive owing to the ``diffusion-controlled
percolation"; (ii) due to the adiabatic localization, the
concentration of conducting electrons decreases as $B^{-1}\ln B$
before the percolation becomes effective, which yields a negative
magnetoresistance $\rho_{xx}(B)\propto B^{-1}\ln B$ for intermediate
$B$; (iii) an exponentially sharp falloff of $\rho_{xx}(B)$ at $B\sim
B_{ad}$ (shown as a vertical jump) separates the diffusive and drift
regimes; (iv) because of the memory effects, the collision time for
scattering by antidots is increased as compared to the Drude value
already in the diffusive regime ($B\ll B_{ad}$), which leads to the
negative magnetoresistance $\rho_{xx}(B)\propto B^{-4}$ for small $B$;
(v) for intermediate $B$, the scattering on antidots stops playing any
role and $\rho_{xx}(B)$ is saturated at a value determined by the
long-range disorder only, whereas at larger fields the
diffusion-controlled percolation gives rise to a positive
magnetoresistance. }
\label{sketch}
\end{figure}

The overall behavior of the MR is illustrated in Fig.~\ref{sketch}.
Different curves correspond to different $n$ for given $l_S$, $l_L$,
and $d$. The characteristic field $B_{ad}$ marks the diffusion-drift
crossover. Figure \ref{sketch} describes the case of not too weak
long-range disorder; specifically, it is assumed that $R_c/l_S\ll 1$
at $B\sim B_{ad}$, which means $l_L/d\ll (l_S/d)^3$.

Having listed the main mechanisms of the MR, let us now summarize our
main results.

We analyze a ``hydrodynamic model" of the chaotic antidot array, i.e.,
tiny antidots scattered with a high density $n\to\infty$ in a smoothly
varying random potential (Sec.~\ref{hydro}). At large $B$,
$\rho_{xx}(B)$ turns out to be a growing power-law function of the
magnetic field. We identified several different regimes of the
behavior of $\rho_{xx}(B)$ [Eqs.~(\ref{e14})--(\ref{e17})], depending on
the parameter $l_S/\sqrt{dl_L}$, where $l_S$ and $l_L$ are the mean
free paths for scattering on antidots and long-range disorder with a
correlation length $d$, respectively. In the limit $B\to\infty$, the
hydrodynamic model universally predicts $\rho_{xx}(B)\propto
B^{10/13}$ [Eq.~(\ref{e17})]. The physics of this divergence is a
percolation of drifting cyclotron orbits limited by scattering on
antidots.

Relaxing the conditions of the hydrodynamic model, we calculate
$\rho_{xx}(B)$ in an antidot array of high but finite density $n$
(Sec.~\ref{finn}). We show that diffusing cyclotron orbits exhibit
intricate dynamics by sticking for a long time to a single
antidot. This leads to a power-law falloff of $\rho_{xx}(B)$ in an
intermediate range of $B$, namely $\rho_{xx}(B)\propto B^{-1}\ln B$
[Eq.~(\ref{e22})]. The small parameter that governs the physics of this
transport regime is $nR_cd$, where $R_c$ is the cyclotron radius. With
further increasing $B$, this mechanism of diffusion is switched off
abruptly, in a critical manner [Eq.~(\ref{e23})].

If the long-range disorder is not too weak, we show that the Drude
regime and the $B^{-1}\ln B$ falloff are connected via an
exponentially fast decrease of $\rho_{xx}(B)$ in a narrow range of the
magnetic field, which is a trace of the adiabatic localization in
smooth disorder (Sec.~\ref{nonad}). In this regime, $\rho_{xx}(B)$ is
determined by the interplay of nonadiabatic transitions and scattering
on antidots [Eq.~(\ref{e25})].

We calculate the scattering time $\tau'_S(B)$ between collisions with
antidots for extended electron trajectories (Sec.~\ref{scatt}), see
Eqs.~(\ref{e27}),(\ref{e30}),(\ref{e33}),(\ref{e38}). At large $B$,
$\tau'_S$ becomes longer than the Drude time. The smaller $n$, the
earlier $\tau'_S$ starts to grow with increasing $B$. In a very dilute
AD array the renormalization of $\tau'_S$ starts already in the
diffusive regime and leads to a falloff of $\rho_{xx}(B)\propto
B^{-4}$ [Eq.~(\ref{e39})].

We solve a scattering problem for a drifting cyclotron orbit colliding
with an antidot at small $\delta/a$ (Sec.~\ref{skipor}). We study
dynamics of cyclotron orbits skipping along the surface of the hard
disc and show that the skipping goes on almost adiabatically, which
results in a strong suppression of transitions between different drift
trajectories [Eq.~(\ref{e53})].

We analyze complex dynamics of skipping cyclotron orbits interacting
with an antidot when the ratio $R_c/d$ is large, which includes
multiple breakaways from the antidot and multiple returns to it,
accompanied by drift along closed trajectories in between
(Sec.~\ref{nonadskip}). We discuss the accumulated effect of these
multiple collisions with a single antidot in terms of the scattering
shift $R_h(B)$ of the guiding center of the cyclotron orbit after it
finally escapes [Eq.~(\ref{e57})]. We also calculate the shift in the
limit of a smoothly varying environment [Eq.~(\ref{e55})].

We calculate the percolative magnetoresistance in the antidot array in
the limit of small $\delta/a$ (Sec.~\ref{rare}). A variety of
different regimes supersede each other with increasing $B$, all of
which are characterized by a power-law behavior of $\rho_{xx}(B)$
[Eqs.~(\ref{e60})--(\ref{e63})]. The asymptotic behavior of
$\rho_{xx}(B)$ in the limit $B\to\infty$ is $\rho_{xx}(B)\propto
B^{1/13}$ [Eq.~(\ref{e63})], which from a practical point of view to
all intents and purposes is indistinguishable from a saturation of the
magnetoresistance.  This behavior is in sharp contrast to the
localization that would develop in the antidot array in the absence of
long-range disorder.

We discuss dynamics of cyclotron orbits which stick to a single
antidot for a long time before hopping to another one (similar to the
$B^{-1}\ln B$ regime in Sec.~\ref{finn}) in the case of a very weak
long-range disorder (Sec.~\ref{short}). Taking this limit eventually
restores the $B^{-1}$ behavior of $\rho_{xx}(B)$ characteristic to the
Lorentz gas.

We present results of numerical simulations (Sec.~\ref{numer}). The
numerical data qualitatively confirm the predictions of the theory.

\acknowledgments We thank M. Heiblum, J. Smet, and V. Umansky for
stimulating discussions. We are grateful to D. Weiss for attracting
our attention to Ref.~\cite{luetjering96}. This work was supported by
SFB 195 and the Schwerpunktprogramm ``Quanten-Hall-Systeme" of the
Deutsche Forschungsgemeinschaft, by INTAS Grants No.\ 97-1342 and
99-1070, and by the German-Israeli Foundation.

\end{multicols}
\end{document}